\documentstyle[prd,aps,psfig,floats]{revtex}

\newcommand{\FRW}{FLRW}
\newcommand{\T}{T}
\newcommand{\ti}{t}

\newcommand{\be}{\begin{equation}}
\newcommand{\ee}{\end{equation}}
\newcommand{\ba}{\begin{eqnarray}}
\newcommand{\ea}{\end{eqnarray}}

\newcommand{\hu}{km~s$^{-1}$~Mpc$^{-1}$}
\newcommand{\cmb}{_{\scriptscriptstyle{CMB}}}
\newcommand{\dab}{D\c{a}browski}

\newcommand{\forget}[1]{\iffalse#1\fi}
\newcommand{\forgetmenot}[1]{\iftrue#1\fi}

\begin{document}

\bibliographystyle{prsty}

\title{Undermining the Cosmological Principle:\\
       Almost Isotropic Observations in Inhomogeneous Cosmologies}

\author{Richard K. Barrett$^{1}$\footnote{Email: \tt{richard@astro.gla.ac.uk}} and
Chris A. Clarkson$^{1,2}$\footnote{Email:
\tt{chris@astro.gla.ac.uk,~clarkson@mathstat.dal.ca}}}

\address{$^1$Department of Physics and Astronomy,~University of Glasgow,~University Avenue,~Glasgow. G12 8QQ.~United
Kingdom.\\ $^2$Department of Mathematics and Statistics,~Dalhousie
University,~Halifax,~Nova Scotia,~Canada,~B3H~3J5.}

\date{\today}

\maketitle

\begin{abstract}

We challenge the widely held belief that the cosmological principle is an
obvious consequence of the observed isotropy of the cosmic microwave
background radiation (CMB), combined with the Copernican principle. We perform
a detailed analysis of a class of inhomogeneous perfect fluid cosmologies
admitting an isotropic radiation field, with a view to assessing their
viability as models of the real universe. These spacetimes are distinguished
from \FRW\ universes by the presence of inhomogeneous pressure, which results
in an acceleration of the fluid (fundamental observers). We examine their
physical, geometrical and observational characteristics \emph{for all observer
positions} in the spacetimes. To this end, we derive \emph{exact, analytic}
expressions for the distance-redshift relations and anisotropies for any
observer, and compare their predictions with available observational
constraints. As far as the authors are
aware, this work represents the first exact analysis of the observational
properties of an inhomogeneous cosmological model for all observer positions.
Considerable attention is devoted to the anisotropy in the CMB. The difficulty
of defining the surface of last scattering in exact, inhomogenous cosmological
models is discussed; several alternative practical definitions are presented,
and one of these is used to estimate the CMB anisotropy for any model. The
isotropy constraints derived from `local' observations (redshift~$\lesssim 1$)
are also considered, qualitatively. A crucial aspect of this work is the
application of the Copernican principle: for a specific model to be acceptable
we demand that it must be consistent with current observational constraints
(especially anisotropy constraints) for all observer locations. The most
important results of the paper are presented as exclusion plots in the 2-D
parameter space of the models. We show that there is a region of parameter
space not ruled out by the constraints we consider and containing models that
are significantly inhomogeneous. It follows immediately from this that the
cosmological principle cannot be assumed to hold on the basis of present
observational constraints.

\end{abstract}


\section{Introduction}

In a previous paper~\cite{cla-bar99} we found the complete class of
irrotational perfect fluid cosmologies admitting an isotropic radiation field
for every observer, and showed that they form a subclass of the Stephani
family of spacetimes that includes the \FRW\ models, but, more importantly,
also includes inhomogeneous cosmologies. Here we study the characteristics of
these inhomogeneous spacetimes from all observer positions in order to assess
their consistency with observations, especially with regard to isotropy
constraints. In particular, we consider whether the \FRW\ models are the only
viable candidates for a cosmological model. A pivotal part of this analysis
is the cosmological principle (CP) and its relation to the Copernican
principle and the observed isotropy of the universe.

Although it is an assumption based purely on philosophy, the Copernican
principle is intuitively very appealing \cite{ellis75,goo}: in order to
reject an inhomogeneous cosmological model on the basis of its conflicting
with the observed isotropy of the universe it is necessary to consider
\emph{all} observer positions and to show that \emph{for most observers in
that spacetime} the anisotropy observed is too large to be compatible with
observations. Actually, in this paper we will, for simplicity, require
consistency with observations for \emph{all} observers~-- our results will
thus be rather stronger than is strictly required by the Copernican Principle.

Having adopted the Copernican principle, the question then
arises as to whether the observed isotropy of the universe, when required to
hold at every point, forces homogeneity, thus validating the CP. Well, the
nearby universe is distinctly lumpy, so it would be difficult to claim there is
isotropy on that basis. However, the CMB is isotropic to one part in 10$^{3}$.
Together with the Ehlers, Geren and Sachs (EGS) theorem~\cite{ehlers-68}, or
rather, the almost EGS theorem of Stoeger, Maartens, and Ellis~\cite{stoeg-95}
this allows us to say that within our past lightcone the universe is almost
\FRW\ (i.e.,~almost homogeneous and isotropic), \emph{provided} the fundamental
observers in the universe follow geodesics (that is, as long as the fundamental
fluid is dust). What the almost EGS theorem achieves is to provide support for
the CP without the need to assume the (near) isotropy of every aspect of the
spacetime: isotropy of the CMB alone is enough to ensure the validity of the CP
(for geodesic observers). Since, however, the Stephani models of
\cite{cla-bar99} do not satisfy the conditions of the almost EGS theorem (the
acceleration is non-zero), its conclusions do not apply; indeed, we know
from~\cite{cla-bar99} that there are inhomogeneous spacetimes admitting
an isotropic CMB (see also~\cite{thesis}).

A number of inhomogeneous or anisotropic cosmological models have been studied
in relation to the CP. The homogeneous but anisotropic Bianchi and
Kantowki-Sachs models \cite{kant-sachs66} (see also \cite{ell98}~\S6, and
references therein) have been investigated with regard to the time evolution
of the anisotropy. It can be shown, for example, that there exist Bianchi
models that spend a significant phase of their evolution in a near-\FRW\
state, even though at early and late times they may be highly anisotropic
(again, see~\S6 of \cite{ell98} and references therein). Most interestingly
for the work presented here, \cite{nils-99} show that there exist Bianchi
models having an almost isotropic CMB but which are actually highly
anisotropic (and have a large Weyl curvature, emphasising that they are
certainly not \FRW), undermining the almost EGS theorem of \cite{stoeg-95}.

Of the inhomogeneous models that arise in cosmological applications, by far
the most common are the Lema\^\i tre-Tolman-Bondi dust
spacetimes~\cite{lem,tolm34,bond47}. These are
used both as global inhomogeneous cosmologies~-- probably the most important
papers being \cite{hel-lak-I84,hel-lak85}, studying geometrical aspects, and
\cite{rindler-sus89,goi-mm97,maar-95,cel99,cel-sch98,mus} investigating
observational aspects~-- and also as models of local, nonlinear perturbations
(over- or under-densities) in an \FRW\ background
\cite{tom96,tomita95,mof-tat95,nakao-95}. See also \cite{kras98} for a review.
Non-central locations are rarely considered in the literature in the analysis
of inhomogeneous cosmologies, owing to the mathematical difficulties that this
usually entails, but \cite{humph-97} made a study of Tolman-Bondi spacetimes
from a non-central location and applied their results to a `Great Attractor'
model. Most other non-central analyses, though, look only at perturbations of
standard \FRW\ models.

However, there has been some consideration of Stephani solutions
\cite{step67a,step67b,kram-80,Kras83,kras97}). These are the
most general conformally flat, perfect fluid solutions~-- and obviously
therefore contain the
\FRW\ models. They differ from \FRW\ models in general because they have
inhomogeneous pressure, which leads to acceleration of the fundamental
observers. \cite{dab-hen98}~fitted a certain subclass of these models to the
first type~Ia supernova (SNIa) data of \cite{perl-97} using a low-order series
expansion of the magnitude-redshift relation for central observers derived in
\cite{dab95}, and found that they were significantly older than the \FRW\
models that fit that data. In \cite{bar-cl99} (see also~\cite{thesis}) we
extend those results, using \emph{exact} distance-redshift relations, and
show that it would always be possible to find an acceptable fit to any data
that could also be fit by an \FRW\ model (for plausible ranges of the \FRW\
parameters $H_0$, $\Omega_0$ and~$\Omega_\Lambda$). The best-fit models were
again consistently at least $1-4$~Gyr older than their \FRW\ counterparts,
with models that provide the best fits to the newer SNIa data \cite{perl-99}
giving an even greater age difference.

One feature of the Stephani models that has been the subject of much debate is
their matter content. The usual perfect-fluid interpretation precludes the
existence of a barotropic equation of state in general (because the density is
homogeneous but the pressure is not), although a significant subclass~-- those
with sufficient symmetry~-- can be provided with a strict thermodynamic scheme
\cite{cla-bar99,bon-col88,sus99}. Individual fluid elements can
behave in a rather exotic manner, having negative pressure, for example
(cf.~Sec.~\ref{energy-cond}). For these reasons, amongst others,
\cite{lorpet86} has claimed that Stephani models are not a viable description
of the universe, but \cite[p.170]{kras97}, argues rather vigorously that this
conclusion is incorrect, as do we. Cosmological models are often ruled out
\emph{a priori} because the matter fails to obey some or all of the energy
conditions \cite{wald,hawk-ell}. However, it has become increasingly difficult
to avoid the conclusion that the expansion of the universe is accelerating,
with the type~Ia supernovae data of \cite{perl-99} being the latest and
strongest evidence for this (see also \cite{kraus98,peeb98}). This suggests
that there is some kind of `negative pressure' driving the expansion of the
universe. For \FRW\ models, this must correspond to an inflationary scenario,
with cosmological constant~$\Lambda>0$, or a matter content of the universe
which is mostly scalar field (`quintessence'~-- see
\cite{framp99,lid99,coble-97,lid-sch98}; see also \cite{gol-ellis98} for a
discussion of the dynamical effects associated with~$\Lambda$). Such matter
inevitably fails to satisfy the strong energy condition. We show in
Sec.~\ref{energy-cond} that the subclass of Stephani models we consider
contains some which satisfy all three energy conditions, and others which do
not. In particular, we suggest in Sec.~\ref{agt0} that the models that would
provide the best fit to the supernova data are models which also break the
strong energy condition.

The main aim of this paper is to discuss the observational effects that arise
in spacetimes admitting an isotropic radiation field when the observer is at
a non-central location (the models we consider have spherical symmetry).
Principally, we are interested in the anisotropy that this will introduce
in observed quantities such as redshift, and we wish to compare these
anisotropies with presently available observational constraints to see
whether it is possible to rule out the inhomogeneous Stephani models
on the basis of their anisotropy. We also examine the anisotropy
of the CMB in these models: although they \emph{admit} an isotropic radiation
field, which is usually implicitly identified with the CMB, it cannot be
assumed that decoupling will actually produce this radiation field in an
inhomogeneous universe. We therefore suggest a number of ways to estimate
the effect of inhomogeneity on decoupling, and determine the resulting CMB
anisotropy. Of course, we can always make any anisotropy as small as we like
by assuming the observer is close to the centre of symmetry. But this is a
very special position, so that such a resolution to the anisotropy problem
would be in conflict with the Copernican principle (although it may be
possible to circumvent this by invoking anthropic arguments~\cite{ellis-78}).
In addition to the anisotropies, though, we also consider the constraints
imposed by measurements of the value of the Hubble constant and the age of
the universe, as well as constraints on the grosser features of the
distance-redshift relations (see Sec.~\ref{size}). While we do not perform
fits to the available magnitude-redshift data here, we assume the results
of \cite{thesis,bar-cl99}, which show that it is possible to fit the
observed SNIa data adequately with the models we consider.

In the following section we describe the spherically symmetric Stephani models
and discuss their physical and geometrical properties, and we present in some
detail the particular two-parameter subclass of the Stephani models that we
will be studying. It will be shown that these models do not have particle
horizons, in contrast to standard \FRW\ models. The energy conditions will
be used to constrain the model parameters, leaving a manageable parameter
set to be investigated further. Then, in Sec.~\ref{noncent}, the
transformation to coordinates centred on any observer will be derived.
In Sec.~\ref{obs-constr} the various distance-redshift relations are
presented for the models we consider, and observations are applied to
constrain the values of the model parameters. The constraints we address are
the value of the Hubble constant, age, the `size' of the spatial sections (the
meaning of which will be explained in Sec.~\ref{size}) and, most importantly,
the anisotropy of the CMB. We show that after applying these constraints there
remains a region of parameter space containing models consistent with all of
the constraints. Furthermore, we demonstrate in Sec.~\ref{inhomog} that many
of the models not excluded by the constraints of Sec.~\ref{obs-constr} are
distinctly inhomogeneous. In Sec.~\ref{local-dipole} we examine the
constraints on the acceleration and the inhomogeneity provided by `local'
(i.e.,~$z\lesssim 1$) observations. In two appendices we derive important
results used in Sec.~\ref{Stephani}.

Readers with families or short attention spans may wish to skip directly to
Sec.~\ref{conclusions}, where the results are summarised and discussed.

\section{The Spherically Symmetric Stephani Models}
\label{Stephani}

Stephani models are the most general conformally flat, expanding, perfect fluid
spacetimes. They have vanishing shear and rotation, but non-zero acceleration
and expansion.   In \cite{cla-bar99} we showed that the irrotational perfect
fluid spacetimes admitting an isotropic radiation field are a subclass of the
Stephani models depending essentially on three free parameters and one free
function of time. Although the general Stephani model has no symmetry at all,
we only consider the class possessing spherical symmetry
(${\mathbf{c}}={\mathbf{0}}$, or ${\mathbf{x}}_0={\mathbf{0}}$, in the
notation of~\cite{cla-bar99}~-- a full analysis of the models
of~\cite{cla-bar99} without spherical symmetry is given in~\cite{thesis}).
The metric in comoving coordinates, from~\cite{cla-bar99}, is
\be
ds^2= \frac{(1+\frac{1}{4}\Delta r^2)^2}{V(r,\ti)^2}\left\{ -c^2 d\ti ^2
+\frac{R(\ti)^2}{(1+\frac{1}{4}\Delta r^2)^2}\left( dr^2+r^2d\Omega^2\right)
\right\}, \label{metric-spec}
\ee
where $c$ is the speed of light, $d\Omega^2 = d\theta^2 + \sin^2\theta d\phi^2$
is the usual angular part of the metric and the function $V=V(r,t)$ is defined
by
\ba
        V(r,t) & = & 1+\frac{1}{4}\kappa (t)r^2: \label{Vdef}\\
        \kappa(\ti) & = & 1-{R}_{,t}(\ti)^2/c^2.\label{kappa-gen}
\ea
$R(t)$ is the scale factor, and $V(r,t)$ is a
generalisation of the \FRW\ spatial curvature factor in isotropic coordinates
(note that in~\cite{cla-bar99} $R(t)$ is included in~$V$).
Since $\kappa$ is a function of~$t$ the spatial curvature can vary from one
spatial section to the next. In fact, it is possible for a closed universe
to evolve into an open universe, or vice versa, in stark contrast to
\FRW\ models~\cite{Kras83}.

We have already restricted ourselves to spherically Stephani models, but we
need to reduce the parameter space further by introducing a form for~$R(t)$
that depends on just a few parameters. To this end, we limit attention to the
two-parameter family derived in Sec.~{IV} of \cite{dab93}. Thus we choose
\be
       R(\ti)  =  c\ti(a\ti+b) \label{Roft}
\ee
where $a$ and~$b$ are the free parameters and
\be
         \Delta \equiv 1-b^2,
\label{Delta}
\ee
so that (as can easily be seen by direct calculation)
\be
                      \kappa(t) = \Delta - \frac{4a}{c}R(t).
\label{kappa}
\ee
We will henceforth refer to these models as D\c{a}browski models. We require
that after the big bang (and before any big crunch) $R>0$, which forces
\be
                   b \ge 0
\label{bpos}
\ee

In (\ref{Roft})--(\ref{Delta}) we retain factors of the speed of light,~$c$,
to facilitate comparison with the references given above and with
observations. The units we will use are as follows:
$[c]=$~km~s$^{-1}$, $r$~is dimensionless,
$R$~is in Mpc and $[\ti]=$~Mpc~s~km$^{-1}=[1/H_0]$, so that $[a]=$\hu$=[H_0]$
and $b$~is dimensionless. Note that these units are slightly different to
those used in \cite{dab95} because the parameters $a$ and~$b$ in that paper
contain a factor of~$c$ (so $[b]=[c]$,~etc.). This explains the appearance
of~$c$ in~(\ref{Roft}).

We will use~$\T$ to denote the coordinate time of a specific epoch of
observation along some observer's worldline (i.e.,~the coordinate age of the
universe), again in Mpc~s~km$^{-1}$, and $\tau$ to denote \emph{proper} time
along a particular flow line. When we state ages they will generally be given
in~Gyr: $\T[{\mathrm{Gyr}}]\approx 978\,\T$[Mpc~s~km$^{-1}$].

Formulae for the expansion and acceleration may be found in~\cite{cla-bar99},
and will be introduced as required, as will formulae for the energy density
and pressure.

\subsection{Geometry}

The metric~(\ref{metric-spec}) is manifestly conformal to a Robertson-Walker
metric with curvature~$\Delta$. If we multiply through by the conformal factor
we see that the spatial sections really are homogeneous and isotropic, but the
\emph{actual} spatial curvature is time-dependent and is given by the curvature
factor $\kappa(\ti)$ in~$V(r,\ti)$. Thus, at any time~$\ti$ whether the
spatial sections are closed, open or flat depends on the sign
of~$\kappa(\ti)$. If, at some point during the evolution of the
universe,~$\ti=-(b\pm 1)/2a$, then the curvature changes sign, as can easily
be seen from (\ref{kappa-gen}) and~(\ref{Roft}). This does not happen in
\FRW\ models, where the spatial curvature,~$k$, is fixed. The distinction
between the time-dependent true geometry ($\kappa$) and the fixed conformal
geometry ($\Delta$) should be borne in mind throughout what follows.

The metric~(\ref{metric-spec}) is not in its most advantageous form. The
conformal geometry of the models is most easily studied by changing from the
stereographic coordinate,~$r$, to the `angle' coordinate,~$\chi$ (see
equation~(5.15) of \cite{hawk-ell}), appropriate to the value of~$\Delta$.
Furthermore, for models with closed spatial sections (which will be our
principal concern here) it is more convenient to choose a radial coordinate that
is better able to reflect the fact that light rays can circle the universe many
times. In such models the spatial surfaces have two centres of symmetry, $r=0$
and~$r=\infty$. Physically, there is nothing extraordinary about the point
$r=\infty$: it is not infinitely far away from the centre, and it is quite
possible for light rays to pass through it. This last point is particularly
important for subsequent discussions, so we make the coordinate change (valid
when~$\Delta >0$, although see below)
\be
r=\frac{2}{\sqrt{\Delta}}\left|\tan \frac{\chi}{2}\right|.\label{r-chi}
\ee
Then $r\rightarrow\infty$ as $\chi\rightarrow\pi$. As a coordinate $\chi$ is
restricted to the range $0\le\chi <\pi$. However, it will prove convenient to
use $\chi$ not just as a coordinate but as a parameter along light rays. In the
latter role its value can increase without bound as the rays circle the universe
many times. Strictly speaking, we should distinguish these two uses, but it
should not lead to confusion. The absolute value is taken in~(\ref{r-chi}) so
that, when $\chi$ increases beyond~$\pi$, $r$ remains positive.
Using~(\ref{r-chi}) in the metric~(\ref{metric-spec}) gives
\be
ds^2 = \frac{1}{W(\chi,\ti)^2} \left\{-c^2d\ti^2+
\frac{R(\ti)^2}{\Delta}(d\chi^2+\sin^2\chi d\Omega^2) \right\},
\label{metric-W}
\ee
where (using (\ref{Vdef} and~\ref{kappa})
\be
W(\chi,\ti)= \cos^2\frac{\chi}{2} V(r(\chi),\ti) = \cos^2\frac{\chi}{2}
           + \frac{\kappa(\ti)}{\Delta} \sin^2\frac{\chi}{2} = 1-\frac{4aR}{c\Delta}\sin^2\frac{\chi}{2}.
\label{confW}
\ee
The conformal factor ($1/W$) is non-singular for all $\chi$ if~$a\le 0$. When
$a>0$ singularities~$W=0$ correspond to spatial and temporal infinity, and
indicate that the universe has `opened up'. As the universe opens up and the
sections become hyperbolic the coordinate~$\chi$ represents a conformal mapping
from a hyperbolic surface onto a sphere: spatial infinity will then correspond
to some finite value of~$\chi<\pi$. For a more detailed explanation of this
see Theorems~4.1--4.4 of \cite{Kras83}.

We can easily calculate the acceleration in these coordinates. It has only a
radial ($\chi$) component:
\be
 \dot{u}_\chi = -c^2 \frac{W,_\chi}{W} = \frac{2ac}{\Delta}\frac{R}{W}\sin\chi.
\label{acceln}
\ee
A simple calculation shows that the acceleration scalar, which we will need
below, is just
\be
  \dot{u} \equiv \left( \dot{u}_a \dot{u}^a \right)^{1/2}
             = \frac{2|a|c}{\sqrt{\Delta}} \sin\chi
\label{accscalar}
\ee
and is therefore time-independent (which, it turns out, is a necessary condition
for a shear-free, irrotational perfect fluid to admit an isotropic radiation
field~-- see \cite{cla-bar99}). Note that the units of~$\dot{u}$ are:
$[\dot{u}]=[c/\ti]=[cH_0]$.

For completeness we mention that when the D\c{a}browski model is conformal to an
\FRW\ spacetime with hyperbolic geometry ($\Delta<0$) the coordinate
transformation is obtained from that just given by replacing trigonometric
functions with their hyperbolic equivalent.

\subsection{Non-Central Observers}
\label{noncent}

Since the purpose of this paper is to provide insight into the observational
characteristics of the D\c{a}browski models for all observer positions, we must
find expressions for the distance-redshift relations and other observable
properties of the models from any point. For a general inhomogeneous metric this
is far from trivial, but the D\c{a}browski models have features that make this
tractable (rather simple, actually). In particular, they are conformally flat.
As has already been noted, in equation~(\ref{metric-W}) the part of the metric
in braces is exactly the form of the \FRW\ metric in `angle' coordinates, so
that the D\c{a}browski models are manifestly conformal to the (homogeneous)
\FRW\ spacetimes. This means that there is a group of transformations acting
transitively on surfaces of constant time that
\emph{preserve the form of the \FRW\ part of the metric} (but not the conformal
factor). If the spatial sections are closed, flat or open (in the conformal
sense, i.e.,~according to the value of~$\Delta$), the transformations are
rotations, translations or `Lorentz transformations', respectively. After such a
transformation the metric will have the form of an \FRW\ metric \emph{centred on
the new point}, multiplied by a modified conformal factor.

As will be shown in Sec.~\ref{energy-cond}, we will be dealing exclusively with
closed models ($\Delta>0$), and so will concentrate on this case. To find the
coordinate transformation to a non-central position, we perform a rotation of
the spatial part of the metric, moving the origin ($\chi=0$) to the point
$\chi=\psi$ ($\psi$~is the observer's position in what follows). In
appendix~\ref{chitrans} we derive this transformation. The old~$\chi$ is given
in terms of the new (primed) coordinates by~(\ref{newchi}). The
conformal factor~(\ref{confW}) then becomes (dropping the primes on $\chi$
and~$\theta$)
\be
W\rightarrow W(\chi,\psi,\theta;\ti)=1-\frac{2aR(\ti)}{c\Delta} (1-
               \cos\psi \cos\chi + \sin\psi \sin\chi \cos\theta),
\label{new-W}
\ee
while the rest of the metric retains its original form (but now in terms of the
new coordinates). This transformation makes the study of our inhomogeneous
models significantly easier, and allows us to find \emph{exact} observational
relations valid for any observer.

\subsection{Lookback Time and the Horizon}
\label{horizon}

If we calculate the lookback time in our models (i.e.,~the time,~$\ti$,
at which a galaxy at some position~$\chi$ emits the light that the observer
sees now at time $\T$), which we can do directly from the
metric~(\ref{metric-W})~-- see appendix~\ref{appen}~-- we get, for any
observer,
\be
t(\chi)=\frac{bT}{(aT+b)\exp{(b\chi /\sqrt{\Delta}) }-aT}.
\label{lookback}
\ee
(Note that this function is continuous through $\chi=\pi$.) Now, $t\rightarrow
0$ if and only if~$\chi\rightarrow\infty$. Thus, the whole of the big-bang
surface is contained within the causal past of \emph{every} observer in the
spacetime and there is no horizon problem for the D\c{a}browski models. This is
in sharp contrast to \FRW\ universes: at early times the particle horizon is
finite and contains only a small part of the big-bang surface, so that widely
separated points can share no common influences. See Figs. 17 and~21 in
\cite{hawk-ell}.

\subsection{Matter Content and Energy Conditions}
\label{energy-cond}

The Stephani models do not have an equation of state in the strict sense, with
the relationship between pressure and energy density being position dependent.
Along each flow line, however, there is a relation of the form $p=p(\mu)$~-- see
\cite{Kras83}. The particular models we are using have an equation of state at
the centre of symmetry (or everywhere in the homogeneous limit~$a\rightarrow 0$)
of the (exotic) form $p=-\frac{1}{3}\mu$. It was shown in~\cite{cla-bar99},
though, that these models do admit a (more general) thermodynamic interpretation
(that is, temperature and entropy can be assigned in such a way
that they are functions only of pressure and energy density). The matter content
of these models with regard to the natural (comoving) velocity field is a
perfect fluid with energy density
\be
                       \frac{8\pi G}{c^2}\mu = \frac{3}{R(t)^2},
\label{density}
\ee
and pressure given by
\ba
     p & = & \mu c^2 \left( \frac{2}{3}
             \frac{V(r,t)}{1+\frac{1}{4}\Delta r^2}-1 \right) \label{pressV}\\
       & = & -\frac{1}{3}\mu c^2
              \left(1+\frac{8aR}{c\Delta}
              \sin^2\frac{\chi}{2}\right)\label{pressW}\\
       & = & -\frac{1}{3}\mu c^2
      \left(1+\sqrt{\frac{24}{\pi
G\mu}}\frac{a}{\Delta}\sin^2\frac{\chi}{2}\right),
\label{pressure}
\ea
where~(\ref{density}) has been used to express the $\ti$-dependence of pressure
in terms of the density. In other words, we have a position-dependent
`equation of state' of the form
$p=-\frac{1}{3}\mu c^2 + \epsilon(\chi)\mu^{1/2}$. The appearance of
$-\frac{1}{3}\mu$ as the dominant contribution (at early times, at
least, when $R$ is small so that, from~(\ref{density}), $\mu$ is large)
immediately suggests a quintessential or scalar field interpretation of the
matter \cite{framp99,lid99,coble-97,lid-sch98}, although it has been shown
\cite{Vilenkin,dab-stel89} that cosmic strings also give rise to this EOS.
Although the energy-momentum tensor of the Stephani models has the perfect fluid
\emph{form}, the interpretation of the matter as an actual fluid is by no means
required; other interpretations may also be valid, and the fact that there is no
true EOS might even suggest a two-component interpretation. For the moment,
though, this is as far as we will go to provide a physical motivation for the
matter in the D\c{a}browski models: in this paper we are only interested in the
observational consequences of the geometry~-- although when we impose the energy
conditions below we show that the matter is certainly not obviously unphysical.

We can see that there are singularities of density and pressure as
$R(\ti)\rightarrow 0$ (i.e.,~at $\ti=0,-b/a$), which correspond to the big bang
and crunch for these models (the metric becomes singular at these points)~-- see
\cite{dab93}. We can also have a \emph{finite-density} singularity, where only
the pressure becomes singular. This happens when $r\rightarrow
2/\sqrt{-\Delta}$. Such infinite pressure is clearly not physical, so we can
reject models with~$\Delta<0$. Models with~$\Delta=0$ are studied in
\cite{dab-hen98}, but it is difficult to compare them directly with $\Delta>0$
models due to the different geometries of the spatial sections so we will not
consider them in this paper, and we are left with $\Delta>0$, i.e.,~$b<1$~--
see~(\ref{Delta}). As we explained in Sec.~\ref{Stephani} the natural
choice of positive $R(\ti)$ after the big bang ensures that $b\ge 0$.

Having calculated the pressure and density of the fluid we can now investigate
its physical viability through the energy conditions. It is more convenient to
use the original stereographic coordinates for this (i.e.,~the expression
(\ref{pressV}) for the pressure), since we wish to consider all values
of~$\Delta$, until we find reasons to the contrary. The weak energy condition
states that $\mu\ge 0$ and $\mu c^2+p\ge 0$, whereas the strong energy condition
requires in addition $\mu c^2+3p\ge 0$ (see, for example, \cite{wald} for a
discussion). The weak energy condition rules out models with finite-density
singularities (i.e.,~with~$\Delta<0$), because such models would contain regions
(for~$r>2/\sqrt{-\Delta}$) where~$\mu c^2+p<0$, as can easily be seen
from~(\ref{pressV}). The weak energy condition also implies that $V\ge 0$, but
this is always true since~$V\rightarrow 0$ only at spatial (and temporal)
infinity (even though the coordinates themselves may be finite). The strong
energy condition, however, implies that $\kappa(\ti)\ge\Delta$ for all~$\ti$.
From~(\ref{kappa}) we can see that this is equivalent to~$a\le 0$
(since~$R>0$), so the models must have a big crunch ($R(\ti)$~is an
`upside-down' quadratic).

The dominant energy condition is more interesting. It states that $|p|\le\mu
c^2$, from which (\ref{density}) and~(\ref{pressure}) immediately give
\[
0\le\frac{1}{3}\frac{V(r,\ti)}{1+\frac{1}{4}\Delta r^2}\le 1.
\]
The left inequality requires only that $\Delta\ge 0$, which rules out
infinities in the pressure. The inequality
on the right says that for all $\ti$ and~$r$
\be
(\kappa(\ti)-3\Delta)r^2\le 8
\label{domc}
\ee
must hold. This condition is always true for~$a\ge 0$ (see~(\ref{kappa})), as
long as~$\Delta\ge 0$. When~$a<0$, $r$~is unbounded for~$\Delta\ge 0$ (because
$\kappa$~is then positive~-- see~(\ref{Vdef})), so the left hand side
of~(\ref{domc}) must always be negative,~$\kappa(\ti)\le 3\Delta$. It is easy to
see that $R(\ti)/c\le -b^2/4a$, so, from~(\ref{kappa}), $\kappa(\ti) \le 1$.
Then $\kappa\le 3\Delta=3(1-b^2)$ for all~$\ti$ provided
\be
          b\le \sqrt{\frac{2}{3}}\approx 0.82.
\label{bdom}
\ee
For~$b$ larger than this the dominant energy condition will be broken at some
time, in regions of the universe at large~$r$. We will not consider further the
intricacies of this. A glance at the exclusion diagrams, Figs.
\ref{exclusion-H50} and~\ref{exclusion-H70}, shows that~(\ref{bdom}) does not
eliminate a significant area of the allowed region. In light of this we will,
for the moment, overlook~(\ref{bdom}) and investigate the properties of all
models with~$0<b<1$.

To summarise: we have used basic physical requirements, such as the occurrence
of a big bang and the avoidance of pressure singularities, along with the energy
conditions to restrict the ranges that the two parameters $a$ and~$b$
(or~$\Delta$) can take. The results are:
\be
    a\le 0,\hskip 0.5cm 0<b<1 \hskip 0.5cm \hbox{(i.e.,~$0<\Delta<1$)}
\label{paramlims}
\ee
(we reject~$b=1$ for simplicity, as explained above, and we refrain from
invoking~(\ref{bdom}) until Sec.~\ref{conclusions}).

\section{Constraining the Model Parameters Using Observations}
\label{obs-constr}

So far we have considered only the `global' physical properties of D\c{a}browski
models, but to really assess their potential viability as cosmological models it
is necessary to confront them with observations. In this section we derive the
distance-redshift relations that form the basis of the classical cosmological
tests and compare them with available observational constraints to see whether
any regions of parameter space are capable of providing a fit. We will impose
constraints on the value of~$H_0$, age, size (the meaning of which will be
explained below) and the anisotropy of the microwave background, leaving the
wealth of data available from galaxy surveys and high-redshift supernovae for
consideration in a future paper; the complexities involved in interpreting such
data and applying it to idealised cosmological models require separate
treatment.

Deriving the observational relations (redshift, angular size or area distance,
luminosity distance and number counts) means relating the coordinates and metric
functions to observable quantities. This requires knowledge of the observer's
motion (4-velocity), which can, strictly speaking, be specified independently of
the background geometry. However, the D\c{a}browski models contain perfect
fluid, so we will identify the observer's motion with the fluid velocity. We are
not obliged to do this, and, given the strange form of the matter, it might be
thought advantageous to instead assume that observers (i.e.,~galaxies)
constitute a dust-like test fluid moving freely through the spacetime whose
geometry is determined by the exotic matter. It will become clear in
Sec.~\ref{CMB} that if we were to make this assumption a large dipole
anisotropy in the CMB would result (although the dipole in~$H_0$ would be
eliminated~-- see~(\ref{specificH0}) with~$\dot{u}=0$) because such a flow
will, in general, have a significant velocity relative to the D\c{a}browski
fluid flow, which, it will turn out, is very nearly in the rest frame of
the CMB everywhere.

First we consider redshift. In general, it is no simple task to find
analytic expressions for the redshift in any cosmological model; derivations
usually rely on symmetries of the spacetime or other simplifying factors to
solve the equations of null geodesics. Here we take advantage of the
conformal flatness of Stephani models (or rather,
of the fact that the D\c{a}browski models are manifestly conformal to \FRW\
spacetimes), although we can also derive the redshift formula as a
time-dilation effect; these procedures are outlined in appendix~\ref{appen}.
Using (\ref{redshift}) and~(\ref{metric-W}) we find
\be
1+z(\psi,\chi,\theta)=\frac{R_0}{W_0}\frac{W(\psi,\chi,\theta;\ti)}{R(\ti)},
\label{redshift-new}
\ee
where $R_0=R(\T)$, $W_0=W(\psi,\T)$ and $\ti$ and~$\chi$ are related by
equation~(\ref{lookback}). Using~(\ref{new-W}) this is
\be
  1+z=\frac{R_0}{W_0}\left\{\frac{1}{R(\ti)}-
\frac{2a}{c\Delta}\left(1-\cos\psi\cos\chi\right) -
\frac{2a}{c\Delta}\sin\psi\sin\chi\cos\theta\right\};
\label{zdipole}
\ee
showing that, for objects at any fixed~$\chi$, the inhomogeneity of universe
manifests itself in the redshift as a pure dipole in angle around the sky
($\cos\theta$~term). This will be important in Sec.~\ref{CMB}.

For metrics with spherical symmetry about the observer the angular size (and
area) distance is given directly from the coefficient in front of the angular
part of the metric, because symmetry ensures that for radial rays $\theta$
and~$\phi$ are constant along the trajectory. For our models we do not have
spherical symmetry about every observer, but the metric is everywhere conformal
to a spherically symmetric metric, as can be seen from~(\ref{metric-W}). Since
null rays are not affected by the conformal factor they also remain at fixed
$\theta$ and~$\phi$, so we can again obtain the angular size distance~$r_A$ from
the coefficient of the angular part of the metric:
\be
r_A(\psi,\chi,\theta)=\frac{R(\ti)}{\sqrt{\Delta}
W(\psi,\chi,\theta;\ti)}|\sin\chi|
\label{areadist-new}
\ee
(again, $\chi$ and~$\ti$ are related by~(\ref{lookback}); the modulus signs
around $\sin\chi$ ensure that~(\ref{areadist-new}) is valid even when~$\chi$ is
treated as a parameter along light rays and takes on values~$>\pi$~--
see~Fig.~\ref{rA-two-b}). We can, for the first time, find the \emph{exact}
angular size-distance relation parametrically by combining
equations~(\ref{areadist-new}) and~(\ref{redshift-new}), which bypasses the
power-series method of \cite{kris-sac66}. This is valid for all sources seen
from \emph{any} observer position.

Luminosity distance~$r_L$ is related to~$r_A$ by the reciprocity theorem:
\be
r_L=(1+z)^2r_A,
\label{recip}
\ee
see \cite{ell98,Mac-Ellis-II}. This then allows the magnitude-redshift
relation to be determined in the usual way: the apparent magnitude~$m$ of
an object of absolute magnitude~$M$ is given in terms of the luminosity
distance by
\be
          m-M-25 = 5\log_{10} r_L.
\label{pogson}
\ee

The task now is to limit $a$, $b$, and~$\T$ using present observational
constraints. A full discussion of each constraint is made in the following
sections, and it is followed by exclusion diagrams showing the regions of
parameter space for which $a$ and~$b$ give a plausible cosmological model for
\emph{all} observer locations in these models.

\subsection{The Hubble Constant}
\label{H0}

The expansion rate of the universe has been measured with reasonable accuracy.
The Hubble constant is believed to lie in the range
$50\lesssim H_0\lesssim 80$~\hu, and we will use these limits to constrain
the D\c{a}browski models. The Hubble \emph{parameter} for these models, which
is related to the volume expansion~$\Theta$, is
independent of position \cite{cla-bar99,kras97}:
\be
H\equiv\frac{\Theta}{3}=\frac{{R}_{,t}(\ti)}{R(\ti)}, \hskip 1.5cm
      H_0=\frac{{R}_{,t}(\T)}{R(\T)}.
\label{Hubble-par}
\ee
We can use this to place constraints on the time at which observations are
made: for our models~$H$ decreases monotonically, so it will only lie in
the observed range of~$H_0$ for some range of~$T$. For any observer with
coordinate age~$\T$, we require
\be
               50\lesssim\frac{2a\T+b}{a\T^2+b\T}\lesssim 80.
\label{min-H0}
\ee
Usually, for simplicity, we will choose a specific value for~$H_0$ (almost
invariably that which produces the `worst case'). Then we can
solve~(\ref{Hubble-par}) for~$T$.

However, when $H_0$ is actually \emph{measured}, it is not necessarily equal
to the expansion rate. What is measured in practice is the lowest order term
in the magnitude-redshift relation, which gives the measured Hubble constant,
$H^m_0$:
\be
H^m_0 = c^2\left.\frac{k^a k^b \nabla_a u_b}{(u_ck^c)^2}\right|_{0},
\label{measured-H0}
\ee
where~$k^a$ denotes the wave-vector of the incoming
photons~\cite{Mac-Ellis-II}. Equivalently, we can consider the gradient of the
redshift-area distance curve at the observer (cf.~\cite{humph-97}).
If we measure the magnitude-redshift relation for objects in some direction,
then, expanding $\nabla_a u_b$ in~(\ref{measured-H0}) in terms of the
kinematical variables rotation~$\omega_{ab}$, shear~$\sigma_{ab}$,
acceleration~$\dot{u}_a$ and expansion~$\Theta$ and using~(\ref{Hubble-par}),
we find (since $\sigma_{ab}=0$ for Stephani models)
\be
H^m_0(\theta) = H_0-\frac{\dot{u}}{c}\cos\theta,
\label{specificH0}
\ee
where $\dot{u}$ is the acceleration scalar and~$\theta$ is the angle between
the acceleration vector and the direction of observation (which is opposite
to the direction in which the photons are travelling). Since the acceleration
is non-zero in the Stephani models there will be a dipole moment in~$H^m_0$.
The size of this in the D\c{a}browski models is given directly
from~(\ref{accscalar}). If this is large in any model we can probably reject
that model because a large dipole moment in~$H_0$ is not observed. However,
very nearby it is difficult to measure~$H_0$ accurately due to peculiar
motions and the discreteness of galaxies. There \emph{is} a dipole moment
in observations of somewhat more distant objects, which is assumed to be due
to the fact that the Local Group is falling into the potential well produced
by Virgo and the Great Attractor. The question is: what upper bound can be
placed on the acceleration by observations? This issue was first raised
in~\cite{cla-bar99}, and will be discussed in more detail in
Sec.~\ref{local-dipole}. For now we simply use~(\ref{Hubble-par}) to
constrain the epoch of observation,~$\T$.

\subsection{The Age of the Universe}
\label{Age}

The original inspiration for \cite{dab-hen98} to study Stephani models was the
potential resolution of the age problem that they provided, which at that time
seemed to be virtually insurmountable within the framework of
\FRW\ models (even when a non-zero cosmological constant was invoked): the high
measured value of~$H_0$ suggested an age,~$\tau_0$, of at most about $11$~Gyr
for an \FRW\ cosmology, whereas globular cluster ages were thought to be up to
$12$-$13$~Gyr~\cite{chab-97}. \cite{dab-hen98,bar-cl99,thesis} showed that,
for the particular Stephani models they considered, this apparent paradox
disappears: the D\c{a}browski models have ages that are consistently 1--4~Gyr
older than their \FRW\ counterparts (for an observer at the centre of
symmetry, at least). However, the age problem has recently been alleviated
by a recalibration of the RR Lyrae distance scale and globular cluster ages
in the light of Hipparcos \cite{chab-97}, which has reduced the globular
cluster ages considerably, to~$\sim 10$~Gyr. The fit is still marginal, but
the new ages are generally accepted as they allow a flat \FRW\ model to fit
the observations provided that $H_0\lesssim 67$~\hu.

We certainly require, then, that at the epoch of observation our models are
older than~$10$~Gyr. However, we will also consider the stronger
constraint~$\tau_0>12$~Gyr, partly to be conservative, but also because the
diagrams for the $12$~Gyr constraint are often clearer.

The age of the universe according to an observer at position~$\psi$ and at
coordinate time~$\T$ is simply the proper time elapsed from the big bang
($\ti=0$):
\be
\tau_0  =  \int_0^{\T}\frac{d\ti}{W(\psi,\ti)}
        =  \frac{\Delta}{4|a|S\sqrt{1-b^2C^2}}
       \ln\left\{
          \frac{\Delta-2a\T S\left(bS+\sqrt{1-b^2C^2}\right)}
               {\Delta-2a\T S\left(bS-\sqrt{1-b^2C^2}\right)}
          \right\}
\label{proper-age}
\ee
(for~$a\ne 0$ and~$\psi\ne 0$, otherwise $W\equiv 1$ and~$\tau_0=\T$), where
$S=\sin(\psi/2)$, $C=\cos(\psi/2)$, and we
take the value of~$\T$ given by the solution of equation~(\ref{Hubble-par}) as
our constraint on the coordinate time for any specific~$H_0$.

\begin{figure}
\centerline{\psfig{file=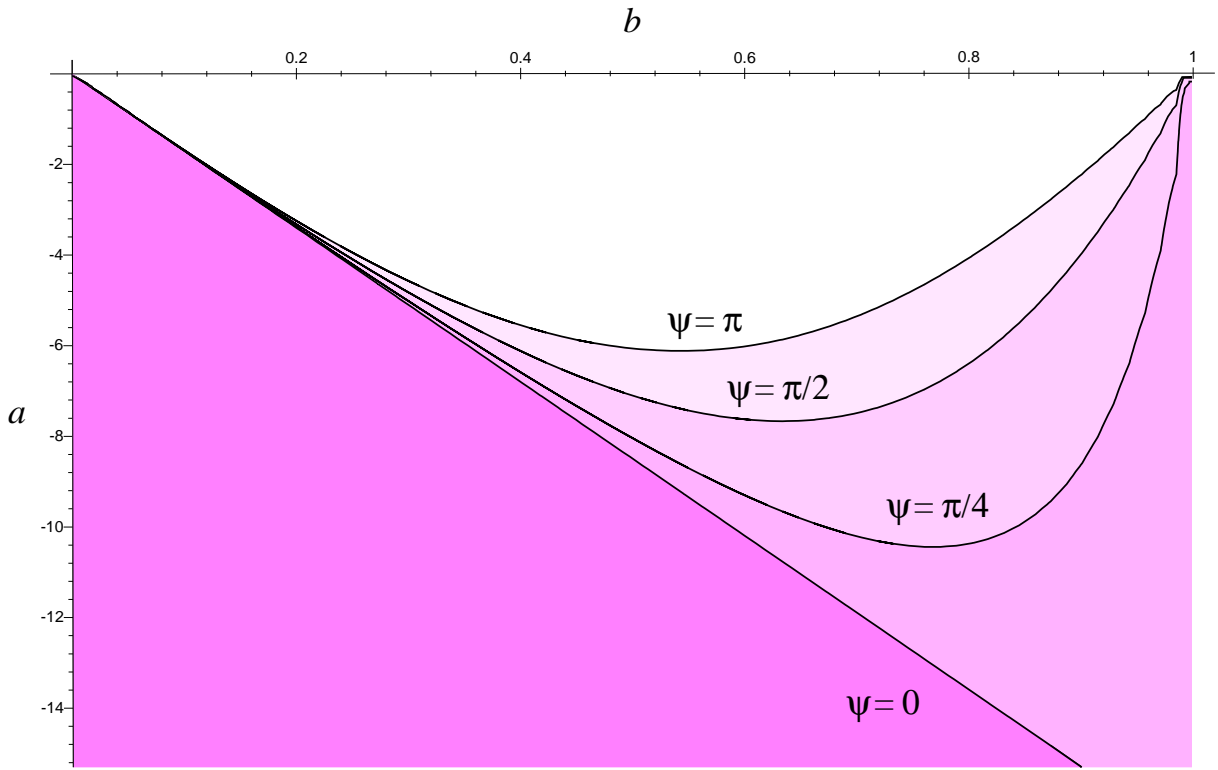,width=16.6cm,angle=90}}
\caption{$\tau_0=12$~Gyr exclusion plot for different observer positions, for
$H_0=60$~\hu. In the shaded areas the models are not old enough. $\psi=\pi$ is
clearly the most restrictive case. } \label{exclusion-age}
\end{figure}

In Fig.~\ref{exclusion-age} the $\tau_0=12$~Gyr contours of the proper-age
function are plotted for~$H_0=60$~\hu\ and for several observer
positions,~$\psi$, showing how the age of an observer varies with~$\psi$ for
different parameters $a$ and~$b$. The shaded regions contain models that are
\emph{less} than $12$~Gyr old for at least one of the observer positions. This
demonstrates that the proper-age of an observer is smallest at the antipodal
centre of symmetry,~$\psi=\pi$. Consequently, we will always use proper-age
at~$\psi=\pi$ to constrain the model parameters. We could weaken this
constraint by requiring only that \emph{most} observers are old enough, which
would allow us to consider instead the age of observers at~$\psi\le\pi/2$ while
still satisfying the Copernican principle (half of the observers would lie in
this region). For simplicity, though, we will not do this here.

\begin{figure}
\centerline{\psfig{file=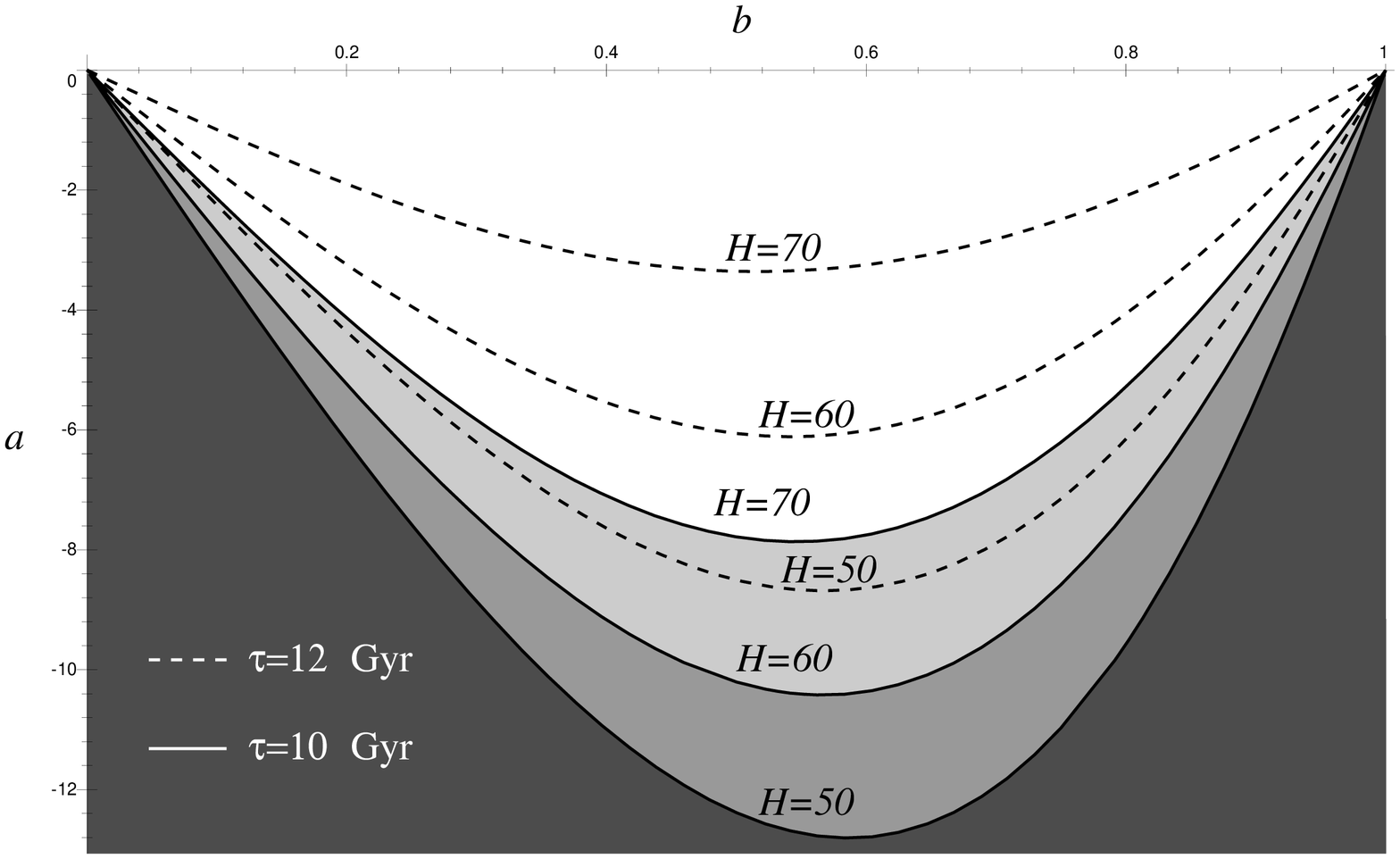,width=16.6cm,angle=90}} \caption{The age
exclusion diagram for various~$H_0$ and proper age~$\tau=10$~Gyr. The shaded
region represents the prohibited area. Also shown as dashed lines are the age
limits for~$\tau=12$~Gyr. (Note that the region excluded for~$H_0=70$~\hu\
contains the excluded regions for lower~$H_0$~-- the progressively darker
shading indicates this.) } \label{exclusion-age-T10}
\end{figure}

Finally, in Fig.~\ref{exclusion-age-T10} we show the age exclusion plot for
the models (based on proper-age at~$\psi=\pi$). We use three values of
$H_0$: $50$, $60$, and $70$~\hu, and a proper-age of~$10$~Gyr (although the
limits for~$12$~Gyr are also indicated). The shaded regions are excluded.
It can be seen that unless we require the universe to be particularly old or
the expansion rate high there is still a significant region of parameter
space that cannot be excluded on the basis of age.

It should be noted here that these plots are meaningless for~$b=1$, because
then~$\Delta=0$ and the model is conformal to an \FRW\ model with \emph{flat}
spatial sections, for which~$\chi$ is not a good coordinate~-- as can easily
be seen from~(\ref{r-chi}).

\subsection{Size and the Distance-Redshift Relation}
\label{size}

When the spatial sections of a cosmological model are closed and there is
no horizon problem light rays may circle the entire universe, perhaps many
times. This is the case for the models we are considering here. What will
the signature of this be in the various distance-redshift relations?
The paths of light rays are determined by the conformal geometry of a
spacetime, and it can be seen from~(\ref{metric-W}) that our models are
conformal to closed, static \FRW\ spacetimes. It follows that light rays
from a point directly opposite
the observer (i.e.,~from the antipode,~$\chi=\pi$) will spread out around the
universe isotropically from the antipode until they pass the `equator'
($\chi=\pi/2$), where they will begin to converge and be focused onto the
observer. As a result, a point source positioned exactly at the antipode will
fill the entire sky when seen by the observer, so that its angular size
distance,~$r_A =$ (physical length/apparent diameter), is zero. Similarly, the
refocusing of light onto a point produces an infinite \emph{flux} at the
observer, and therefore the luminosity distance,~$r_L$, is also zero
($m\sim\log_{10}r_L=-\infty$). It is obvious that whenever the light rays
travel through a parameter distance~$\chi$ that is an exact multiple of~$\pi$,
$r_A=r_L=0$: this is reflected by the factor of $\sin\chi$
in~(\ref{areadist-new}).

\begin{figure}
\centerline{\psfig{file=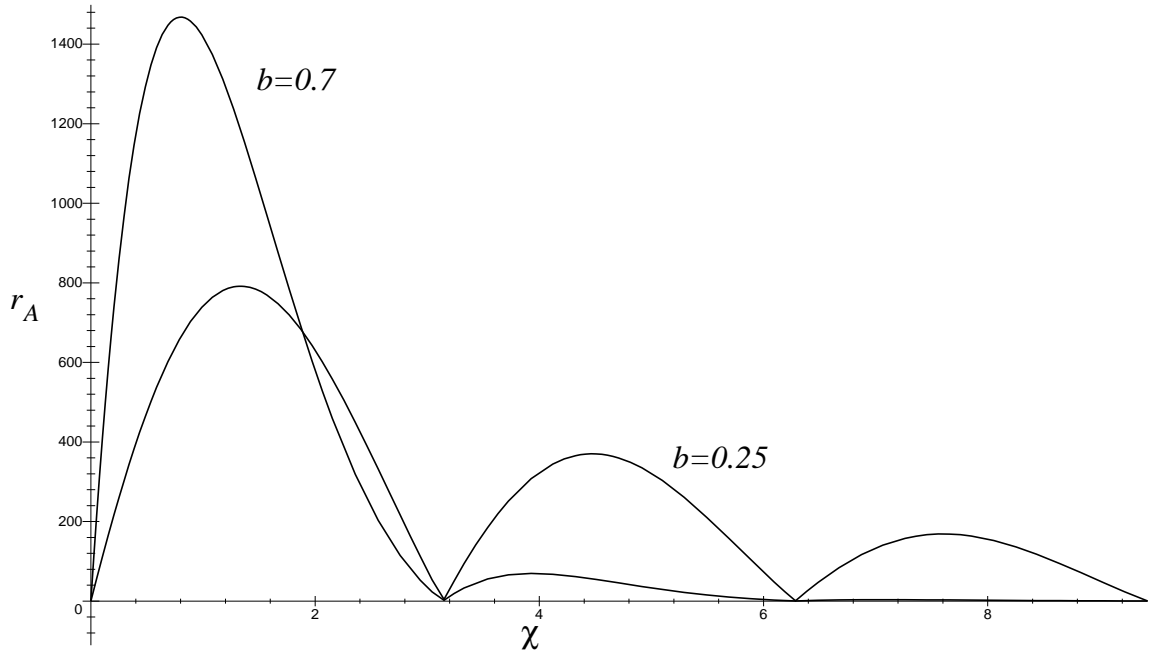,width=16.6cm,angle=90}} \caption{Area
distance from the centre as a function of~$\chi$ for two values of~$b$.
$\T=15$~Gyr,~$a=-1$.} \label{rA-two-b}
\end{figure}

\begin{figure}
\centerline{\psfig{file=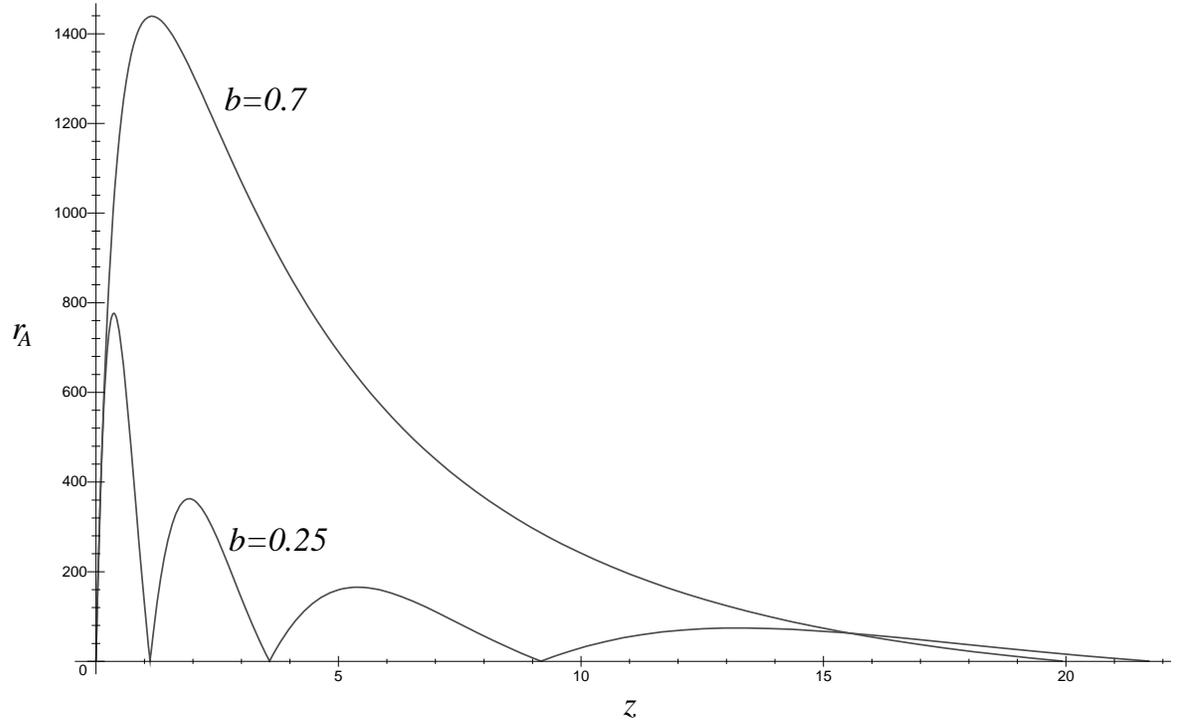,width=16.6cm,angle=90}} \caption{Area
distance from the centre as a function of redshift for the same parameters as
Fig.~\ref{rA-two-b}. For small~$b$ the angular size distance oscillates far too
rapidly.} \label{rA-z-two-b}
\end{figure}

\begin{figure}
\centerline{\psfig{file=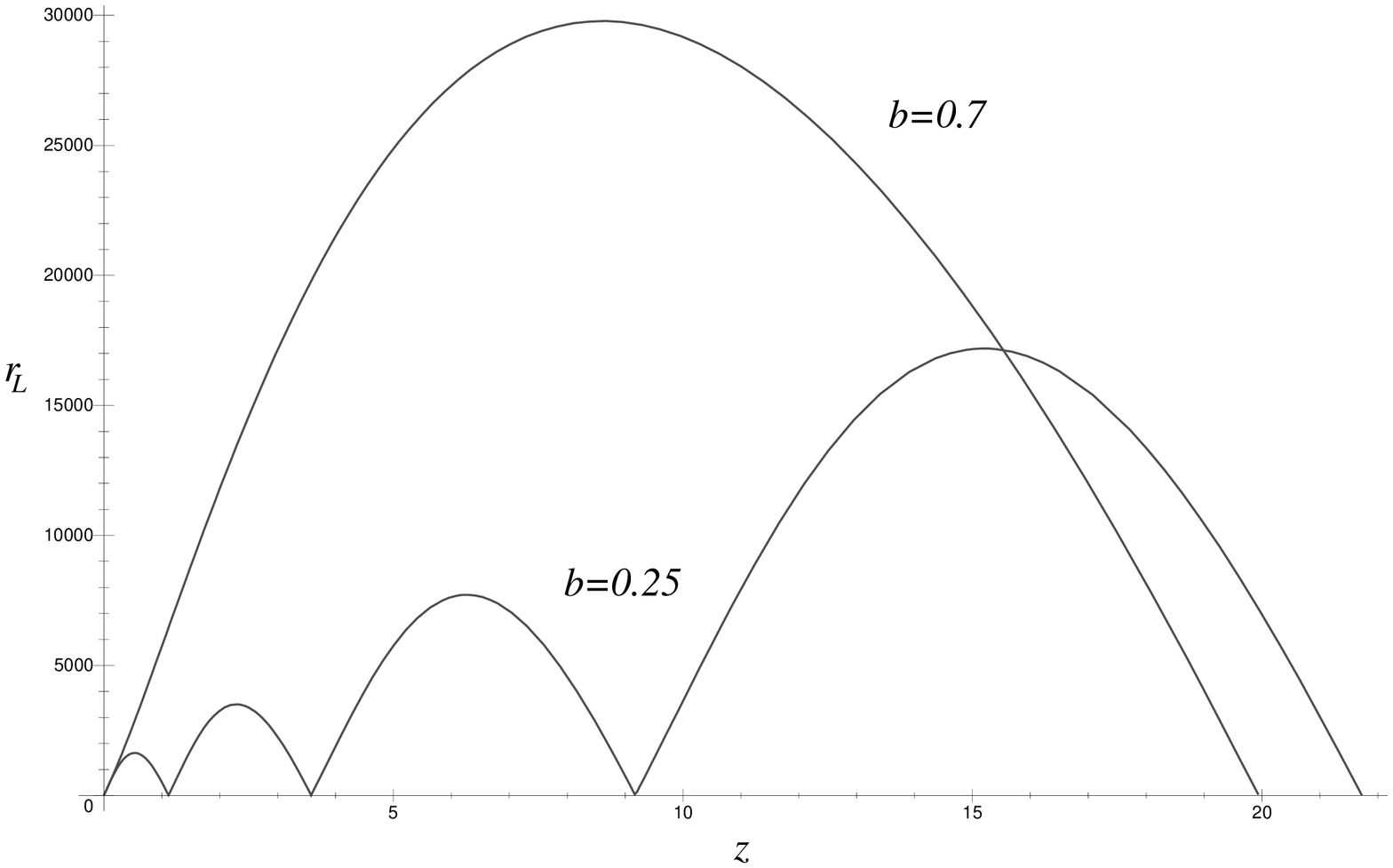,width=16.6cm,angle=90}} \caption{Luminosity
distance from the centre as a function of redshift for the same parameters as
Figs. \ref{rA-two-b} and~\ref{rA-z-two-b}.} \label{rL-z-two-b}
\end{figure}

This effect can be seen clearly in Figs.~\ref{rA-two-b}--\ref{rL-z-two-b},
where we show the two principal measures of distance as they vary with
coordinate distance $\chi$ or redshift. Viewed as a function of~$\chi$, in
Fig.~\ref{rA-two-b}, the zeros of the angular diameter distance occur
at multiples of~$\pi$ for all model parameters. Looked at in terms of redshift,
though (Fig.~\ref{rA-z-two-b}), it is clear that for small~$b$ the zeros are
much closer together than for larger~$b$, with the first zero occurring
at~$z\approx 1$ for~$b=0.25$. Figure~\ref{rL-z-two-b} shows the luminosity
distance-redshift relation, for comparison. These effects are not as unusual
as they look, and can be found also in \FRW\ geometries for models with
positive~$\Lambda$~-- see~\S4.6.1 in \cite{ell98} and references therein.


Can we rule out such apparently aberrant behaviour? Theories of structure
formation are fairly well developed (see \cite{bert-97} for a thorough
discussion and references), and the evolution of galaxies and the star
formation rate~(SFR), while not accurately known, are at least qualitatively
understood. In particular, the SFR, which is very important for determining
the luminosity of distant, young galaxies, is believed to fall off
beyond~$z\sim 2$ \cite{madau96,sadat98}. As a result, one could argue that
there will be relatively few bright objects beyond some redshift~$z_{SF}$
that corresponds to the epoch at which galaxies `turned on' and the SFR
began to increase significantly. This would mean that the zeros in the
distance-redshift relations would be essentially unobservable if they
occurred at redshifts larger than~$z_{SF}$ because there would be no
luminous objects to be seen magnified in the sky, whereas if the zeros
occurred at lower redshifts than~$z_{SF}$ one could reasonably argue that
there ought to be some signature of this in the observations. Since galaxies
have only been observed (in the Hubble Deep Field, for example) with
redshifts up to~$z\sim 5$, and quasars have only been seen out to a similar
redshift, we take $z_{SF}=5$.

\begin{figure}
\centerline{\psfig{file=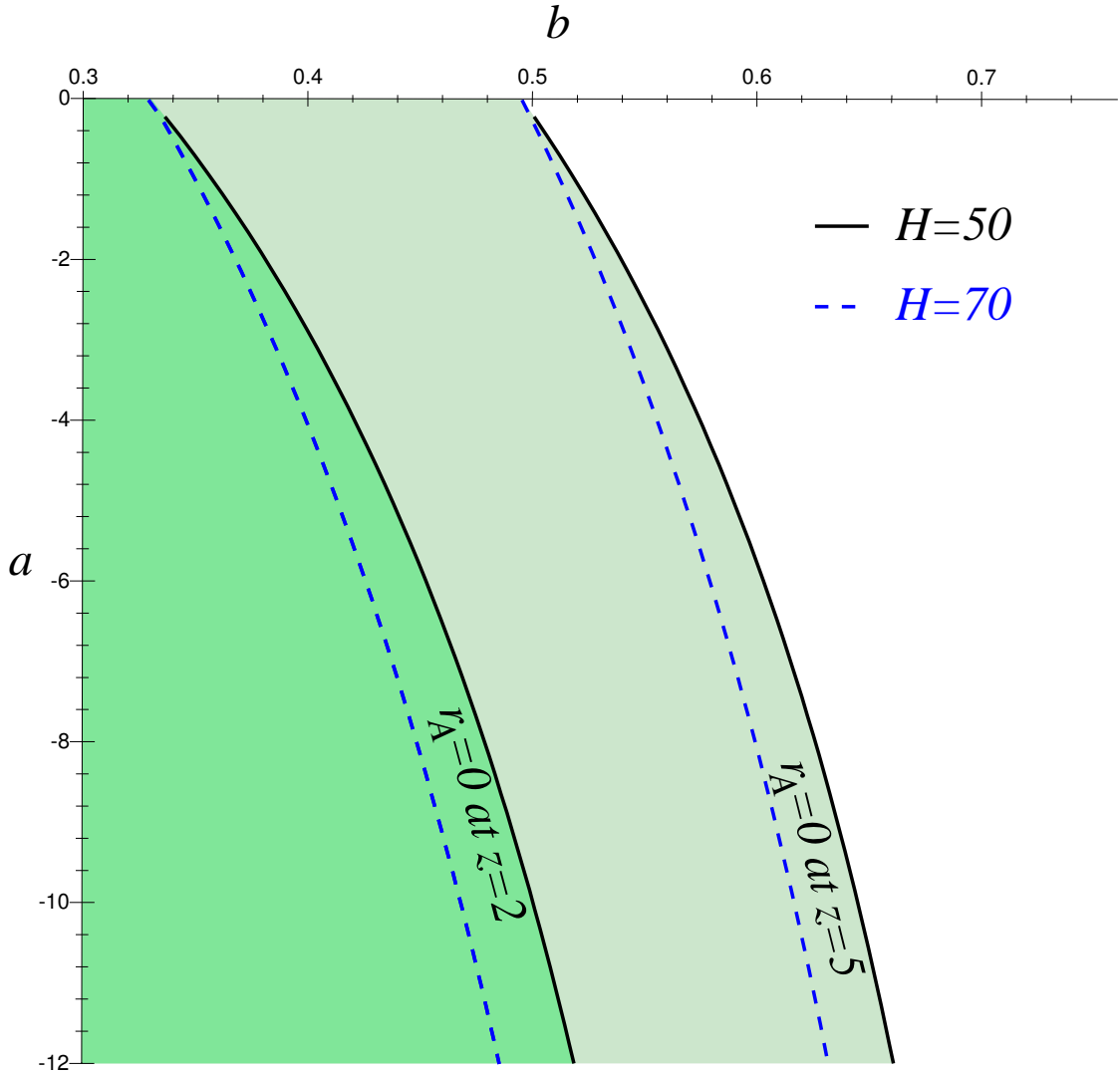,width=16.6cm,angle=90}} \caption{Exclusion
plot obtained by requiring that the first zeros of~$r_A(z)$ occur at~$z>z_\pi$,
for $z_\pi=2$ and~$z_\pi=5$. The shaded regions are excluded. Curves are given
for $H_0=50$ and $H_0=70$~\hu.} \label{rA-exclusion}
\end{figure}

\begin{figure}
\centerline{\psfig{file=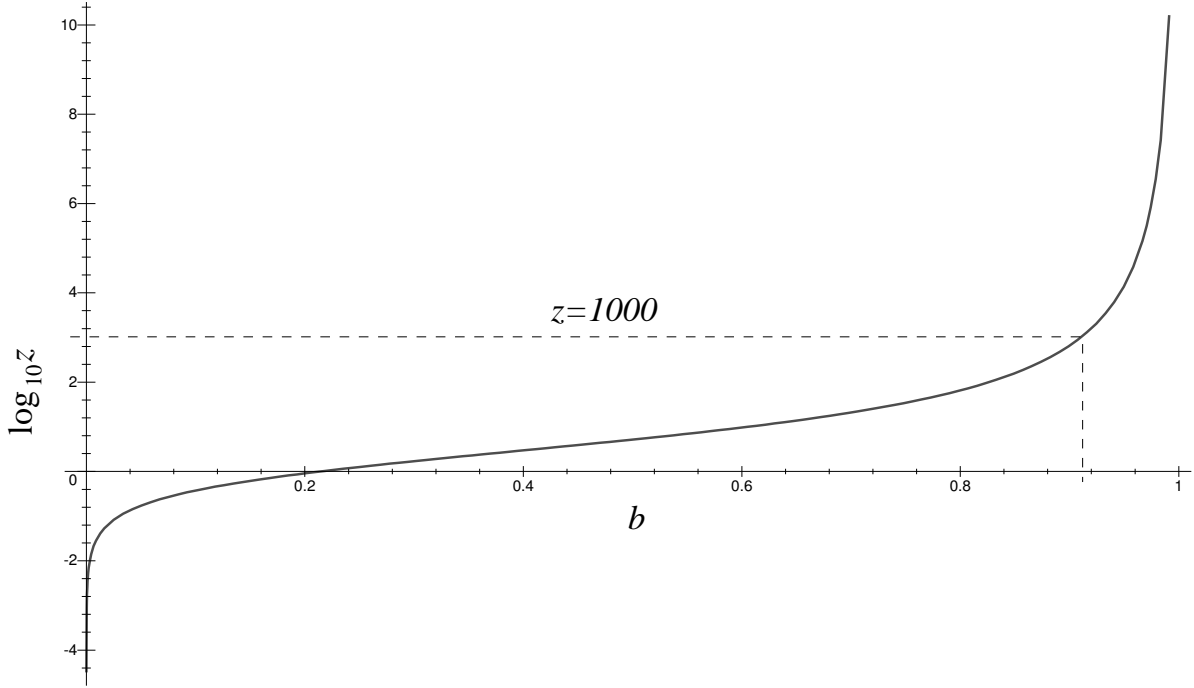,width=16.6cm,angle=90}} \caption{Logarithmic
plot of the redshift at which the first zero of~$r_A$ occurs as a function
of~$b$ for $H_0=50$~\hu\ and~$a=-1$. For the first zero to occur at~$z>1000$,
so that recombination occurs `nearer to us' than the antipode~--
i.e.,~at~$\chi<\pi$~-- requires $b\protect\gtrsim 0.9$.} \label{logz-b}
\end{figure}

The constraints imposed by larger~$z_{SF}$ can be inferred from Figs.
\ref{rA-exclusion} and~\ref{logz-b}. (For example, if~$z_{SF}\sim 30$, the
redshift at which it is suggested that the very first luminous objects may
appear~\cite{loeb99}, Fig.~\ref{logz-b}, gives $b\gtrsim 0.7$ for~$a=-1$.)
There are, of course, a number of factors that rather cloud these arguments,
in particular the possible effects of extinction, which are not well
known, as well as the fact that the number of objects `near' to the antipode
will be small, because the spatial volume is diminished as a result of the
geometry. Interestingly, number counts in the HDF show a dip at~$z\sim 2$
\cite{wang98,saw-97} which would occur in any model for which the antipode
is at this redshift. However, this dip is generally believed to be
spurious~\cite{cooray99}.

If these arguments are not completely convincing, then at a simpler level the
fact that the \emph{observed} magnitude-redshift relation is known accurately
out to~$z\sim 1$ from type~Ia supernovae \cite{perl-99}, and is certainly not
dipping down, allows us to say that there is no zero of luminosity distance
below~$z\sim 2$, say. We therefore also consider the constraint that results
from requiring that there are no zeros below~$z=2$.

We wish, then, to constrain the parameters of our models by rejecting any models
for which the first zero in the distance-redshift relations occurs at~$z\le
z_\pi$, where $z_\pi=2$ or~$z_\pi = z_{SF}=5$. Using~(\ref{redshift-new}), this
means
\be
  1+z(\chi=\pi)=\frac{R_0}{W_0}\frac{W(\psi,\pi,\ti_\pi)}{R(\ti_\pi)}>1+z_\pi,
\label{redshift-rA0}
\ee
where $R_0=R(\T)$, $W_0=W(\psi,\T)$ (the conformal factor at the observer) and
$\ti_\pi$ denotes the lookback time~(\ref{lookback}) at~$\chi=\pi$. Again we
determine the epoch of observation (i.e.,~the observer's coordinate time~$\T$)
using~(\ref{Hubble-par}). The solution of~(\ref{redshift-rA0}) for $a$ and~$b$
is shown in Fig.~\ref{rA-exclusion} as an exclusion diagram. The effect of this
constraint is to rule out small values of~$b$, for any~$a$. This is a reflection
of the fact that, loosely speaking, $b$~measures the `size' of the universe: at
early times the scale factor goes as~$b\ti$, so that when $b$~is small the
spatial sections are small, light rays don't take long to travel from antipode
to observer, the scale factor changes relatively little during this time and the
redshift of the antipode (which is dominated by $R_0/R(t_\pi)$ as in \FRW\
models) is small.

As a coda to this section we consider the effect of demanding that the first
zero of~$r_A$ is effectively unobservable as a result of being `hidden' behind
the CMB. Figure~\ref{logz-b} shows how the redshift of the first zero of
$r_A(z)$ varies with~$b$. If, instead of choosing~$z_\pi=z_{SF}$ as our primary
constraint, we want the first zero of $r_A(z)$ to happen at a redshift large
enough for the universe to be opaque (i.e.,~before decoupling), then
Fig.~\ref{logz-b} shows that $b$ must be quite close to unity. This figure also
allows the extent to which values of~$b$ are excluded for any~$z_\pi$ to be
estimated.

\subsection{The Microwave Background Anisotropy}
\label{CMB}

The CMB is observed today to be a blackbody at a temperature of
$T_0=2.734\pm0.01$K, with a dipole moment of $T_1=3.343\pm0.016\times10^{-3}$K
and quadrupole moment as large as $T_2=2.8\times10^{-5}$K (see \cite{part97}
for details and references). It was emitted at a time when the radiation was
no longer hot enough to keep Hydrogen ionised, causing it to decouple from
matter, which happens at $T_{\mathrm{dec}}\sim 3000$K. Idealised cosmological
models do not have realistic thermodynamics (that is, they do not, in general,
accurately describe the thermodynamic evolution of the gas and radiation
mixture that fills the real universe). In
\FRW\ models the epoch at which decoupling occurs is simply defined to be that
corresponding to the redshift necessary to shift the temperature at decoupling
to the observed mean temperature of the CMB,~$T_0$. From the redshift relation
applied to the temperature of a blackbody ($T$~will be used to denote
temperature in this section),
\be
                        T_{\mathrm{obs}}=\frac{T_{\mathrm{dec}}}{1+z},
\label{CMB-temp}
\ee
we infer that the CMB is formed at a redshift~$z\approx 1000$. This definition
is fine for homogeneous models, leading to a consistent definition of the time
of decoupling for every observer at the same cosmic time, but raises an
interesting point for the inhomogeneous D\c{a}browski models, because the
redshift depends on both the observer's position,~$\psi$, and the angle around
the sky,~$\theta$. If we simply define the redshift of the CMB at any point to
satisfy~(\ref{CMB-temp}) with $T_{\mathrm{obs}}=T_0$ then, by definition, we
obtain a perfectly isotropic CMB for that observer, but we must choose a
different emitting surface for each different observer. Such an observer-based
definition of the CMB surface is clearly unsatisfactory.

In fact, we know already from the results of \cite{cla-bar99} that the models
we consider \emph{admit} an isotropic radiation field for every observer,
which is implicitly identified with the CMB. However, this identification
overlooks any physics underlying the production of the CMB. When, as with
the original EGS theorem, the models under consideration turn out to be
\FRW\ this is acceptable, since it can be assumed that the homogeneity
applies to the production of the CMB: at some moment of cosmic time
decoupling occurs everywhere throughout the universe. Unfortunately, when
the models admitting an isotropic radiation field are inhomogeneous this
is no longer acceptable, and some consideration must be given to the
production of the CMB and how this may be affected by the different
conditions at different places in the universe. To give detailed
consideration to the physics of decoupling in the non-standard D\c{a}browski
cosmologies would take us beyond the scope of this paper, so instead we
consider several alternative, pragmatic definitions of what is meant by the
CMB surface, which, while not fully capturing the physics of decoupling,
at least allow the influence of inhomogeneity at the time of decoupling to be
estimated (it is important in these definitions to distinguish between the
dominant exotic matter that is responsible for the geometry of the
D\c{a}browski models~-- see Sec.~\ref{energy-cond}~-- and the putative
`real' baryonic gas that decouples):
\begin{enumerate}
\item \label{EGSdef} If we avoid all consideration of the physics of
decoupling, we \emph{could} simply assume that it happens at such an early
time that we could define the CMB to be free-streaming radiation `emitted
at the big bang', as is effectively assumed in the EGS and almost~EGS theorems
\cite{ehlers-68,stoeg-95}. This only really makes sense if there is some
natural definition of the radiation field at early times. For example, if the
model is homogeneous and isotropic at early times, we can define a
homogeneous and isotropic radiation field. Our models have exactly this
property of homogeneity at early times (as can be seen from~(\ref{confW}),
$W\rightarrow 1$ as~$\ti\rightarrow 0$). For our models this definition will
result in a perfectly isotropic CMB for every observer \cite{cla-bar99}
(although its observed temperature will be position dependent);
\item \label{rhodef} Ideally we would like to define decoupling in terms of
the thermodynamics of the baryonic gas (at a fixed temperature $T\sim p/\rho$,
say). However, the D\c{a}browski matter is not an ideal gas and its pressure
and density are not that of the real baryonic gas, so the utility of this
definition here is limited;
\item In general inhomogeneous models we could choose a fixed value of
some physical quantity such as density which would allow us to estimate the
degree of inhomogeneity~-- for the D\c{a}browski models this is equivalent
to~\ref{costdef} because the density is homogeneous on cosmic time surfaces;
\item \label{costdef} We could choose a surface of constant cosmic
time,~$\ti=\ti\cmb$~-- since the D\c{a}browski models possess a cosmic time
coordinate with respect to which only the pressure is inhomogeneous this is
a natural extension of the \FRW\ definition;
\item \label{ptdef} We could take a surface of constant \emph{proper} time
(based on the assumption of some common evolution for the ideal gas component
at different positions)~-- see equation~(\ref{proper-age}).
\end{enumerate}
We will not consider~\ref{EGSdef} here for the reasons outlined above. Also,
the homogeneity of D\c{a}browski models at early times means that for
small~$\ti$ the proper-age is virtually identical to the coordinate time
(equation~(\ref{proper-age}) with~$W\approx 1$). It turns out that for times
of observation that reproduce the observed~$H_0$ any reasonable definition
of the CMB surface puts it at an early time, which means
definition~\ref{ptdef} is virtually identical to~\ref{costdef}. We will
therefore define the CMB according to~\ref{costdef} in this section.
This amounts to assuming that the early homogeneity allows for
homogeneous physics on the CMB surface just as in \FRW\ models, and that the
small inhomogeneity that is present affects only the redshift of points on the
CMB surface. Note that the anisotropy of the CMB that arises from this
definition of the CMB surface does not conflict with the results of
\cite{cla-bar99}: a radiation field that is isotropic for every observer may
still be defined, but any realistic process that gives rise to the CMB must
reflect the inhomogeneity of the universe at the time of decoupling; the CMB
anisotropy we derive in this section results essentially from the (small)
inhomogeneity on the CMB surface.

It still remains, though, to decide exactly \emph{which} surface of constant
cosmic time the CMB originates from. Consider, for an observer at
position~$\psi$, the temperature distribution on the sky that the CMB would
have if it were emitted from the surface~$\ti=\ti\cmb$ (related by the lookback
time formula~(\ref{lookback}) to some distance~$\chi\cmb$):
\be
T_{\mathrm{obs}}(\psi,\chi\cmb,\theta)=
           \frac{T_{\mathrm{dec}}}{1+z(\psi,\chi\cmb,\theta)}.
\ee
Equation~(\ref{zdipole}) shows that we can write
\[
  1+z(\psi,\chi\cmb,\theta) = 1+z_0(\psi,\chi\cmb) +
                      z_1(\psi,\chi\cmb)\cos\theta
\]
where
\ba
   1+z_0(\psi,\chi\cmb) & = & \frac{R_0}{W_0}\left[\frac{1}{R\cmb}-
\frac{2a}{c\Delta}\left(1-\cos\psi\cos\chi\cmb\right)\right],\nonumber \\
   z_1(\psi,\chi\cmb) & = & -\frac{2a}{c\Delta}\frac{R_0}{W_0}
   \sin\psi\sin\chi\cmb
                        = \frac{\dot{u}(\psi)}{c^2}
                        \frac{R_0}{\sqrt{\Delta}W_0}\sin\chi\cmb
\label{z1}
\ea
(using~(\ref{accscalar}) in the last equality and assuming~$a\le 0$). The mean
redshift of the CMB surface is~$z_0$; $z_1$~gives rise to an anisotropy in the
CMB. We can therefore define the location of the CMB surface to be
the~$\ti\cmb$ (or~$\chi\cmb$) that gives a mean redshift of~1000. That is,
$\chi\cmb$ is the solution of
\be
         z_0(\psi,\chi\cmb) = 1000
\label{zcmbdef}
\ee
for any observer position~$\psi$.

Having found~$\chi\cmb$ we can evaluate the anisotropy in the temperature of the
CMB. Since $T_{\mathrm{obs}}$ depends on the reciprocal of~$1+z$ the dipole
moment in~$z$ will give rise to higher multipoles when expanded as a binomial
series:
\be
T_{\mathrm{obs}}(\theta)=\frac{T_{\mathrm{dec}}}{1+z_0}\left[1-
\frac{z_1}{1+z_0}\cos\theta
    +\left(\frac{z_1}{1+z_0}\right)^2 \cos^2\theta
    + O(\cos^3\theta)\right],
\ee
that is,
\[
\frac{\delta T(\theta)}{T}=-\frac{z_1}{1+z_0}\cos\theta
    +\left(\frac{z_1}{1+z_0}\right)^2 \cos^2\theta
    + O(\cos^3\theta).
\]
The dipole moment of the CMB temperature is then (using~(\ref{zcmbdef}))
\be
                \delta_1 = \frac{z_1}{1+z_0} \approx 10^{-3}z_1.
\label{Tdipole}
\ee

Measurements of the CMB can now be used to constrain the model parameters. We
at least require that the dipole moment should be no larger than the observed
dipole anisotropy,~$|\delta_1|<T_1/T_0\approx 10^{-3}$ (i.e.,~$|z_1|<1$). If
this is satisfied for any model then it is clear that the quadrupole and
higher multipole moments will all be~$\lesssim 10^{-6}$~-- certainly no larger
than their observed values. In fact, such a constraint on~$z_1$ is very weak,
leaving vast tracts of parameter space entirely untouched. Moreover, there
are very good reasons for believing that there is a significant contribution
to the observed dipole moment from the peculiar velocity of the Local Group
as a result of infall towards the Great Attractor~\cite{lynd-bell88}, which
can be measured with moderate accuracy using galaxy surveys, and seems to be
consistent with the motion of the Local Group with respect to the
CMB~\cite{schmoldt-99,ries-95}, although see~\cite{lau-pos91}. Actually, it is
not beyond the bounds of possibility that genuinely inhomogeneous background
models such as those we
consider here could mimic the grosser features of the local anisotropies (in
particular the GA induced dipole effect~-- see Secs.~\ref{H0}
and~\ref{local-dipole}) as well as the CMB dipole. The `real' universe would
then be a perturbation of this inhomogeneous background. Viewed in this way,
local observations (galaxy surveys,~etc.) would reveal the effects of the
large-scale inhomogeneity and also contain information about smaller-scale
perturbations (peculiar velocities and the density contrast). That is, the
peculiar velocity field and the density contrast that are inferred assuming an
\FRW\ background would actually be thought of as containing one part that
reflects the difference between the inhomogenous and \FRW\ background models
(and therefore contains the dipole effect referred to above, for example) and
another part that is the `true' perturbation. (The distinction between these
two components is a fine one: ultimately it reflects the difference between
linear and fully nonlinear perturbations of \FRW\ models.) We will consider
this in a future paper \cite{cla-rau99}. However, we choose here to reject
iconoclasm in favour of the more conservative viewpoint that most of the
observed dipole \emph{is} due to the peculiar motions induced by local
inhomogeneities, but that there remains some leeway~-- up to 10\% of the
observed dipole~-- due to observational uncertainties, for there to be a
purely cosmological contribution to the CMB dipole. Then the largest dipole
moment that we can accept from our models is~$|\delta_1|<10^{-4}$, or
\be
                           |z_1|<0.1.
\label{maxzdipole}
\ee

\begin{figure}
\centerline{\psfig{file=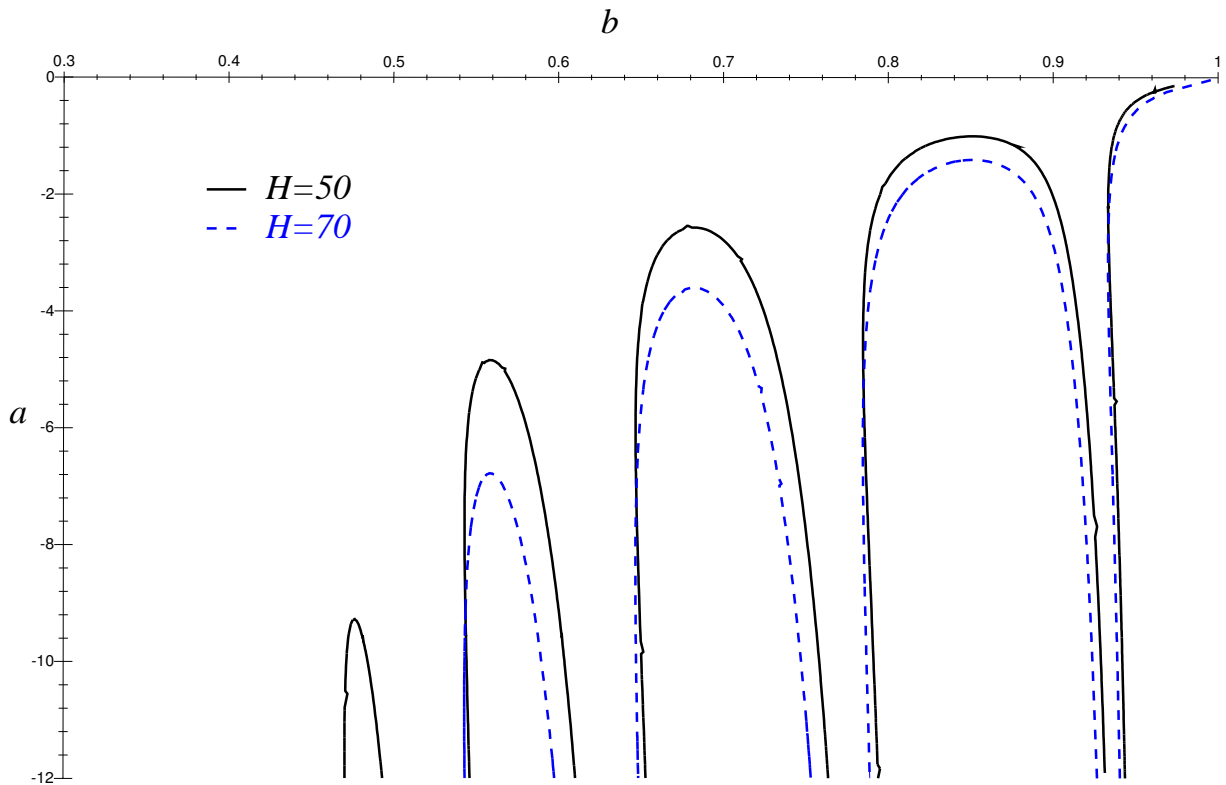,width=16.6cm,angle=90}} \caption{The
exclusion diagram from CMB anisotropies for an observer at $\psi=\pi/2$. The
curves for $H_0=50$ and $70$~\hu\ are shown. As $H_0$ increases, the `fingers'
move down to more negative~$a$. The allowed region lies between and above the
fingers.} \label{cmb-exclusion}
\end{figure}

Given any model parameters and some observer position we adopt the following
procedure. First we use~(\ref{Hubble-par}) to determine the epoch of
observation for some~$H_0$, as usual, then we solve for~$\chi\cmb$
using~(\ref{zcmbdef}). Having found all the parameters we need to
determine~$z_1$ we simply check~(\ref{maxzdipole}) to see whether the model,
or at least that observer position, must be rejected. In practice we can
simply solve~(\ref{maxzdipole}) as an equality to obtain $a$ as a function
of~$b$ at the boundary of the allowed region, and this is what is shown in
Fig.~\ref{cmb-exclusion} (for $\psi=\pi/2$), where it can be seen that a
low value of $H_0$ constrains our models most~-- in contrast to the age
constraint. This is because~$H_0$ decreases monotonically with time, so
small~$H_0$ corresponds to a later time of observation and therefore a
later time for the CMB surface, which means that~$W$ has evolved to become
more inhomogeneous. We choose $\psi=\pi/2$ because, as is clear from
(\ref{acceln}) and~(\ref{z1}), the anisotropy is generally worst there, so
if a model is rejected at~$\psi=\pi/2$ it will be unacceptable everywhere.
Again, therefore, for models not excluded in Fig.~\ref{cmb-exclusion}
detection of a CMB anisotropy of the magnitude that we observe would be
typical, and the Copernican principle need not be abandoned for these models
to be viable.

The finger-like excluded regions in Fig.~\ref{cmb-exclusion} appear because
for different model parameters~$\chi\cmb$ takes on different values, and for
some parameters this value is very close to (a multiple of)~$\pi$, so that
the entire CMB seen by an observer in such a model is emitted from virtually
a single point (the antipode). Since redshift depends only on the
relative conformal factors at emitter and observer for our conformally flat
models, the CMB must be almost exactly uniform however inhomogeneous the
model (i.e.,~whatever the value of~$a$).

\subsection{The Local Dipole Anisotropy}
\label{local-dipole}

Although we are not in a position to use real observations to constrain the
dipoles that would be detected in observations of the `local' universe (in
galaxy surveys, for example, where~$z\lesssim 0.01$, or with type~Ia supernova
data, for which~$z\lesssim 1$), we can at least consider these effects
qualitatively. It will turn out that for the models we consider observations
of the dipole variation in the distance of objects at a given redshift (or the
dipole in~$z$ at a given distance) directly constrain the acceleration of the
fundamental observers. In fact, it is clear from~(\ref{specificH0}) that this
is the case for \emph{any} model (at low redshift at least), so that
measurements of these dipole moments at low~$z$ permit the
acceleration~$\dot{u}^a$ to be measured. It is then only a question of the
tightness of the constraints imposed by real observations, and the extent to
which the acceleration signal can be separated out from other components in
the peculiar velocity field. This we consider in \cite{cla-rau99}. Note that
the acceleration dipole grows with distance at low~$z$, as can be seen
from~(\ref{specificH0}), so that it is distinct from the dipole resulting
from GA infall, which amounts to an overall Lorentz boost by a constant
factor and is therefore independent of distance (see~\cite{cla-bar99}). It
should be noted that this is only true when objects more
distant than the GA are observed; closer in the dipole structure due to GA
infall is more complicated, and at very small distances there should be no
dipole variation in~$H^m_0$ (to first order), because~$\dot{u}=0$ in the
standard interpretation (although there will be a quadrupole component
due to shear).

In this section we adopt the null hypothesis that the universe \emph{is} well
described by a (perturbed) \FRW\ model (i.e.,~that observations are at least
consistent with~$\dot{u}=0$), and determine whether, and under what
circumstances, local observations may provide a tighter constraint on the
D\c{a}browski models than the CMB (Fig.~\ref{cmb-exclusion}).

From (\ref{zdipole}) and~(\ref{accscalar}) the redshift dipole for objects at
any coordinate radius~$\chi$ from an observer at~$\psi$ (in~(\ref{accscalar})
$\chi$ is the observer position relative to the centre of symmetry, which
we now denote~$\psi$) is seen to be
\begin{equation}
|z_1(\psi,\chi)| =
\frac{\dot{u}(\psi)}{c^2}\frac{R_0}{\sqrt{\Delta}W_0}|\sin\chi|
\label{z1gen1}
\end{equation}
(cf.~equation \ref{z1}). For small~$z$ this just corresponds to the
$H^m_0$~dipole in~(\ref{specificH0}), since~$z=(H^m_0/c)r_{\mathrm{prop}}$ (at
low~$z$ the distance measures $r_A$ and $r_L$ are the same as proper
distance~$r_{\mathrm{prop}}$). It is clear that the redshift anisotropy for
objects at some distance from the observer is closely related to the
acceleration~$\dot{u}$ of the fundamental observers and the dipole moment in the
measured~$H_0^m$ in~(\ref{specificH0}). Defining~$\delta H=\dot{u}/c$, the
dipole in redshift for objects at a given~$\chi$ becomes
\be
|z_1(\psi,\chi)| = \frac{\delta H}{c}\,\frac{R_0}{\sqrt{\Delta}W_0}|\sin\chi|.
\label{z1gen2}
\ee

If we assume that the time of observation,~$\T$, is fairly close to~$\ti=0$
(as is generally the case for the values of~$H_0$ we allow), then models
with~$a<0$ have $\T\ll -b/2a$ (the time at which the scale factor reaches its
maximum value and $H_0=0$) and $W_0\approx 1$, so that $R_0/c\approx b\T$ and
${R_{,t}}_0/c\approx b$. Equation~(\ref{z1gen2}) then becomes, with the help
of~(\ref{Hubble-par}),
\be
|z_1(\psi,\chi)| \approx \frac{{R_{,t}}_0}{c\sqrt{\Delta}}|\sin\chi| \,
                                 \frac{\delta H}{H_0}
                 \approx \frac{b}{\sqrt{\Delta}}|\sin\chi|\frac{\delta H}{H_0}.
\label{z1gen3}
\ee
Note that dependence on distance from the observer only arises through
the~$\sin\chi$ factor, so for objects at any given distance from the observer
\be
        |z_1(\psi,\chi)| \lesssim \frac{b}{\sqrt{\Delta}}\frac{\delta H}{H_0}.
\label{z1gen4}
\ee
In principle, therefore, local observations of redshift anisotropies determine
the~$H_0$-dipole, and therefore the acceleration~$\dot{u}$, through
(\ref{z1gen1}) and~(\ref{z1gen2}). Conversely, any measurement of the
acceleration or Hubble dipole can be used to constrain the dipole
anisotropy~$z_1$ at all redshifts through (\ref{z1gen3}) or~(\ref{z1gen4}).

Before considering the constraints imposed by local observations, we
compare~(\ref{z1gen4}) with the CMB constraint~(\ref{maxzdipole}) imposed in
Sec.~\ref{CMB}. If we assume that the bound in~(\ref{maxzdipole}) is reached,
i.e.,~that the CMB dipole due to the inhomogeneity of the D\c{a}browski model
we are considering is~$|z_1|=0.1$, then it follows from~(\ref{z1gen4}) that
\be
             \frac{\delta H}{H_0}\ge 0.1 \frac{\sqrt{\Delta}}{b}
\label{dHCMB}
\ee
($\sqrt{\Delta}/b$~is a decreasing function of~$b$, so that this constraint is
stronger for smaller~$b$). That is, if the universe was well described by a
D\c{a}browski model and the cosmological CMB dipole was measured to be as
large as the bound specified in~(\ref{maxzdipole}), then the local
$H_0$-dipole must be larger than~$0.1\sqrt{\Delta}/b$. (For example, a
variation in~$H_0$ of less than~20\% can only arise if $\sqrt{\Delta}/b\le 2$,
or~$b\ge 1/\sqrt{5}\approx 0.45$; for a variation of less than~10\% we must
have~$b\ge 1/\sqrt{2}\approx 0.71$.)
Since the CMB anisotropy constraint~(\ref{maxzdipole}) allows models
with~$b\gtrsim 0.5$ (see Fig.~\ref{exclusion-H50}), corresponding to a
variation in $H_0$ around the sky of at least 17\% for models on the boundary
of the allowed region in Fig.~\ref{exclusion-H50} where~$|z_1|=0.1$, if
present observations show that the $H_0$-dipole is less than 17\% some of
the models allowed by~(\ref{maxzdipole}) will be excluded by these local
observations. Note, though, that~(\ref{dHCMB}) is considerably
stronger than is really required for most model parameters, owing to the fact
that the~$\sin\chi$ factor in~(\ref{z1gen3}) was neglected. This amounts to
adopting as the CMB constraint the envelope of the fingers in
Figs.~\ref{cmb-exclusion}, \ref{exclusion-H50} and~\ref{exclusion-H70}, which
would obviously overlook large areas of parameter space that should really be
allowed.

Although it would seem that the easiest way to constrain the acceleration in
the D\c{a}browski models would be to measure the local $H_0$-dipole, it is not
really possible to measure~$\delta H/H_0$ accurately~\cite{cla-rau99}.
Moreover, there is known to be a dipole in observations of galaxies at
somewhat larger redshifts ($z\sim 0.01$), which is usually interpreted as
the effect of infall of the Local Group towards the Great Attractor
\cite{lynd-bell88,schmoldt-99}. This infall is manifest as a systematic
relative motion of the Local Group with respect to distant galaxies at a
velocity~$v\approx 600$~km~s$^{-1}$, coinciding with the Local Group motion
relative to the CMB frame. In order not to conflict with these observations we
at least require that at redshifts~$z_0\approx 0.01$ the dipole moment due to
the D\c{a}browski acceleration is no larger than the observed dipole:
\[
                   c z_1 \lesssim v = 600\hbox{~km~s$^{-1}$}.
\]
Then, since at low redshift the linear Hubble law is valid, we have
\[
     cz_0=H_0r_{\mathrm{prop}},\quad   cz_1 = \delta H r_{\mathrm{prop}},
\]
and so
\be
\frac{\delta H}{H_0} =        \frac{cz_1}{cz_0}
                     \lesssim \frac{600\hbox{~km~s$^{-1}$}}{0.01c} = 0.2.
\label{GAdipole}
\ee
That is, we can allow at most a 20\% variation in~$H_0$ around the sky. This
is comparable to the 17\% variation required by the CMB limit derived above,
which means that~(\ref{GAdipole}) does not provide a stronger constraint on
the D\c{a}browski model parameters than~(\ref{maxzdipole}). Interestingly,
\cite{rub-76a,rub-76b} estimated the dipole in~$H_0$ to be about~20\% (the
Rubin-Ford effect). However, this measurement has since been discredited
\cite{jam-91}, being the result of selection effects.

If the $H_0$-dipole could actually be measured, or at least bounded, then
the time independence of the acceleration scalar in~(\ref{accscalar}) makes
it a simple matter to use this to constrain the model parameters $a$ and~$b$.
From (\ref{specificH0}) and~(\ref{accscalar}) (with~$\chi=\psi$)
\[
         \frac{\delta H}{H_0} \equiv \frac{\dot{u}}{cH_0}
                         = \frac{2|a|}{\sqrt{\Delta}H_0}\sin\psi.
\]
If $\delta H/H_0\le\gamma$, say, and we assume this holds for all observer
positions~$\psi$, in accordance with the Copernican principle, then
\[
             |a| \le \frac12 \gamma H_0 \sqrt{\Delta}.
\]

It should be borne in mind that throughout the preceding discussion we have
only considered the modulus of the dipole, not its direction. It is clear
from~(\ref{z1}) that the local dipoles may be in the same direction as the CMB
dipole or in the opposite direction, depending on the sign of~$\sin\chi\cmb$.
What is more, the variation of the dipole with distance is controlled entirely
by~$\sin\chi$, so the dipole will change sign whenever~$\chi$ is a multiple
of~$\pi$. The alignment of the CMB and local dipoles is usually taken to be a
strong sign that both result from the peculiar motion of the Local Group (due
largely to GA infall). However, it is clear that for the models we consider
these dipoles will also always be aligned (or anti-aligned).

\subsection{The Combined Exclusion Diagrams}

\begin{figure}
\centerline{\psfig{file=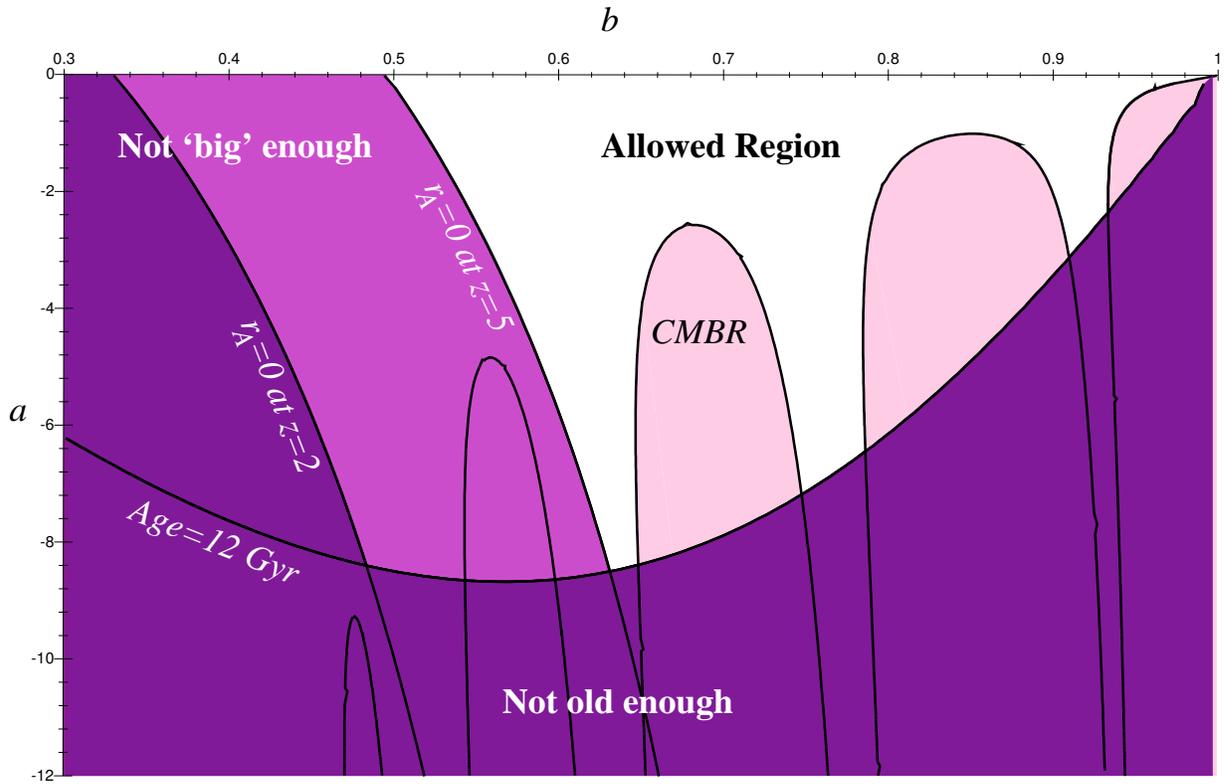,width=16.6cm,angle=90}} \caption{The
complete exclusion diagram for all the observational constraints studied (age,
size and the CMB anisotropy), for~$H_0=50$~\hu. We have taken the $12$~Gyr age
constraint, to be conservative. The dominant energy condition should be added
to these constraints: it eliminates models with~$b>0.82$
(equation~(\ref{bdom})).} \label{exclusion-H50}
\end{figure}

\begin{figure}
\centerline{\psfig{file=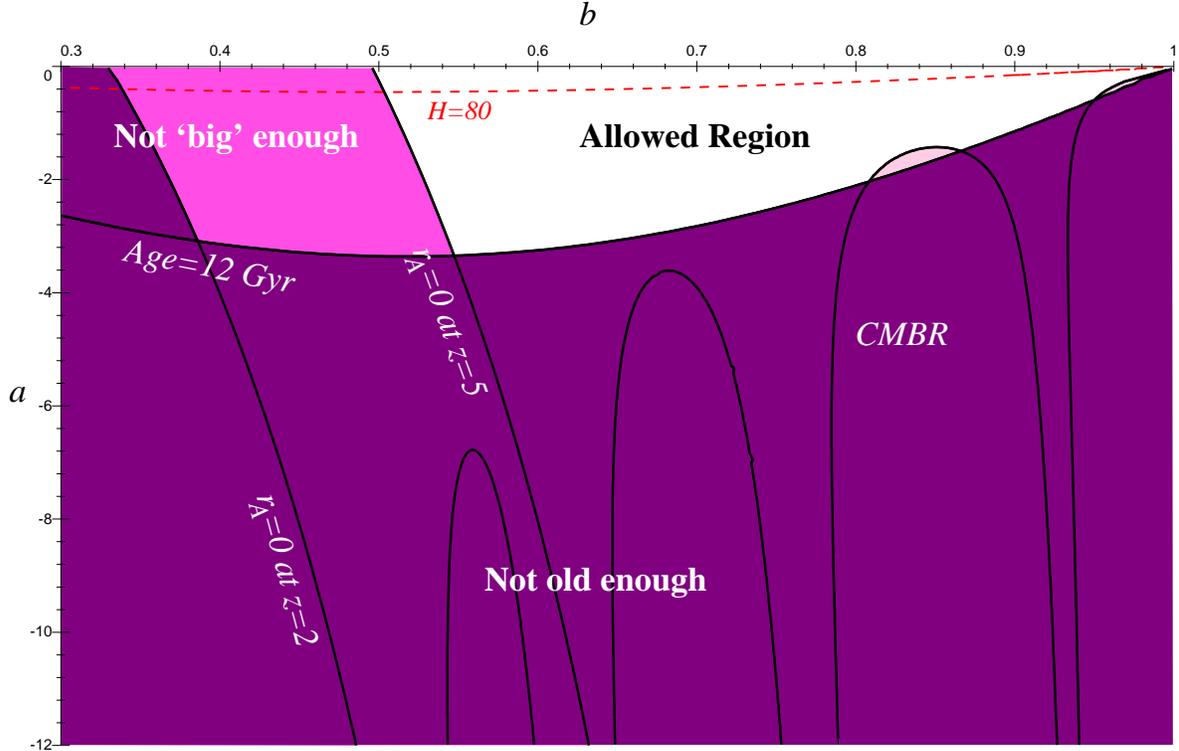,width=16.6cm,angle=90}} \caption{\small As
in Fig.~\ref{exclusion-H50}, but for~$H_0=70$~\hu. The age constraint
for~$H_0=80$~\hu\ is also shown as a dashed line (close to the $b$-axis):
high~$H_0$ means a low age.} \label{exclusion-H70}
\end{figure}

When we combine all of the constraints derived in this section (Figs.
\ref{exclusion-H50} and~\ref{exclusion-H70}) we can see that for~$H_0=50$~\hu\
the strongest constraint comes from the CMB, with age placing somewhat weaker
limits on the allowable degree of inhomogeneity (which is measured largely by
the size of~$a$~-- see Sec.~\ref{inhomog}). The `size' restriction of
Sec.~\ref{size} eliminates quite a large region of parameter space for
small~$b$, but this is not really a constraint on the inhomogeneity, which is
our principal concern.

Perhaps rather surprisingly, given the results of \cite{dab-hen98,bar-cl99},
the strongest constraint for larger~$H_0$ really comes from the age. As can
be seen from the exclusion plot for~$H_0=70$~\hu, Fig.~\ref{exclusion-H70},
the CMB constraint pokes out in places to eliminate certain regions, and the
size constraint cuts off low values of~$b$, but age does most of the dirty
work. It can also be seen that for $H_0=80$~\hu\ age imposes a very strong
constraint on the models (dashed line in Fig.~\ref{exclusion-H70}): models
with high~$H_0$ must be very nearly homogeneous, or they are too young.
However, if we relaxed the age constraint to $10$~Gyr the CMB anisotropy
would be the dominant limitation for most values of~$H_0$ in the currently
fashionable range ($50\lesssim H_0\lesssim 80$~\hu).

We should not forget, at this point, to reintroduce the
restriction~(\ref{bdom}) from the dominant energy condition, which rules
out high~$b$. This is not shown on the diagrams, in order to avoid clutter.
Most of the models eliminated by this constraint have already been ruled
out by the age or CMB constraints, and models that are rejected solely by
the dominant energy condition are not hugely inhomogeneous (see the next
section).

\section{The Size of the Inhomogeneity}
\label{inhomog}

While we have considered many different aspects of the D\c{a}browski models,
what we have not done is to assess the extent to which the models that are not
excluded are inhomogeneous. It is obvious from the exclusion plots,
Figs.~\ref{exclusion-H50} and~\ref{exclusion-H70}, that the homogeneous
D\c{a}browski models (those with~$a=0$) are the `most acceptable', in that all
the constraints favour small~$a$. This should not be surprising, as far as
anisotropy constraints are concerned, at least. What is not clear is whether
the allowed region only contains models that are very nearly homogeneous.
We will show that it does not.

The most natural way to assess the degree of inhomogeneity of the models is
to examine the variation of the pressure over surfaces of constant cosmic
time. It can be seen from~(\ref{pressW}) that the extremes of pressure occur
at $\chi=0$ and~$\chi=\pi$, so we define the inhomogeneity factor~$\Pi$ to
be the relative pressure difference between the two poles:
\be
          \Pi = \left|\frac{p(\pi)-p(0)}{p(0)}\right| = \frac{8|a|R}{c\Delta}.
\label{Pidef}
\ee
If~$\Pi\gtrsim 1$ then it is reasonable to say that the models are truly
inhomogeneous, whereas if~$\Pi\ll 1$ they are obviously nearly \FRW. Note,
though, that~$\Pi$ depends on the cosmic time surface under consideration: for
small~$\ti$, $\Pi\approx 0$, and~$\Pi$ reaches its maximum at~$\ti=-b/2a$
(see~(\ref{Roft})), at which time
\[
               \Pi = \frac{2b^2}{1-b^2} = 2\frac{1-\Delta}{\Delta}
\]
so that the models are significantly inhomogeneous ($\Pi\gtrsim 1$) when
\be
          b\gtrsim \frac{1}{\sqrt{3}}\approx 0.58.
\label{binhom}
\ee
Most of the allowed models in Figs.~\ref{exclusion-H50} and~\ref{exclusion-H70}
satisfy~(\ref{binhom})~-- models with smaller~$b$ have already been eliminated
by the size constraint in Sec.~\ref{size}.

This is not really a fair reflection of the inhomogeneity of the models at the
times of observation that are relevant here, though, because the~$H_0$
constraint~(\ref{Hubble-par}) generally ensures that the epoch of observation
is quite early on in the evolution of the universe when the scale factor is
somewhat smaller than its maximum size. To evaluate the impact of this,
consider two specific examples. From the allowed region of
Fig.~\ref{exclusion-H50} choose the model at $a=-7$,~$b=0.75$.
For~$H_0=50$~\hu\ the solution of~(\ref{Hubble-par}) is
$\T\approx 0.016$~Mpc~s~km$^{-1}$, which gives~$\Pi=1.33$. For the model at
$a=-8$, $b=0.64$ we get $\T = (3-\sqrt{5})/50 \approx 0.015$~Mpc~s~km$^{-1}$
and~$\Pi=0.86$, which is close enough to~$1$. These models are certainly
not `close to \FRW'.

Nevertheless, it could be said that the models are not massively
inhomogeneous, and that their degree of inhomogeneity only reflects the
looseness of the constraints applied. This is not so: even for models that
are only inhomogeneous at the 10\% level ($\Pi=0.1$), the CMB
anisotropy,~$\delta_1<10^{-4}$, is at least \emph{three orders of magnitude
smaller} than the inhomogeneity that generates it. This certainly conflicts
with the spirit of the almost EGS theorem of~\cite{stoeg-95} which says that
small CMB anisotropies indicate correspondingly small perturbations from
homogeneity. Of course, we know from \cite{cla-bar99} that the D\c{a}browski
models all \emph{admit} a perfectly isotropic radiation field; the fact that
the CMB is not precisely isotropic here is a result of our more
sophisticated definition of the CMB surface.

\section{A Note on Models with $\lowercase{a}>0$}
\label{agt0}

In this paper we have only considered the effects of the models
with~$a\le 0$ in order to satisfy the strong energy condition. The
type~Ia supernovae data strongly implies that the expansion rate of the
universe is increasing ($q_0<0$) which, in the $\Lambda=0$ \FRW\ case
at least, can only occur when the strong energy condition fails, as may be
seen from the Raychaudhuri equation. This is not necessarily true when
acceleration is present however~\cite{thesis,ell98}, but we may derive
the following condition for the D\c{a}browski models to have~$q_0<0$ at
some time~$\ti$:
\[
\frac{a}{\Delta}\left[\Delta + \left((2a\ti+b)^2+b^2\right)\sin^2\frac{\psi}{2}
\right]>0.
\]
The term in brackets is always positive (unless~$\Delta<0$, which we do not
consider), so that $q_0<0$ if and only if~$a>0$. In fact, this is intuitively
clear from the quadratic form of~$R(\ti)$: the accelerated expansion implied
by the SNIa data can only be produced when~$R(\ti)$ is an `upright'
quadratic~($a>0$).

By analogy with \FRW\ models, we may define a Hubble normalised density
parameter
\[
         \Omega(\ti) = \frac{24\pi G\mu }{\Theta^2} = (2a\ti+b)^{-2},
\]
which we may use to get a rough idea of how these models behave in comparison
with \FRW\ models. The present day value of the density parameter is roughly
in the range
$
          0.3 \lesssim \Omega_0 \lesssim 1.
$
For the D\c{a}browski models, we note that in the limit $\ti\rightarrow 0$ we
have $\Omega \rightarrow {1}/{b^2}$, and $\Omega_{,t} \rightarrow
-{4a}/{b^3}$. We know that $b$~is positive in order that the scale factor is
positive, which implies that $\Omega_{,t}$~is negative at the big bang only
if~$a>0$. However, we also see that as $\ti\rightarrow 0$, then
$\Omega$~approaches some value larger than~$1$ if~$b<1$ (which is required
if we reject models with finite-density singularities). This means that if
we desire $\Omega\lesssim 1$ today, then we need~$a>0$, in contradiction
with the strong energy condition.

When $a>0$, the quantitative restrictions from the age, size and CMB
anisotropies are not strongly affected; indeed, in some ways the outlook is
better. As is shown in \cite{thesis,bar-cl99}, the proper age of an observer
in an~$a>0$ model is considerably larger than their compatriot in a
negative~$a$ model, for the same value of~$H_0$; the `size' of the universe
(i.e.,~the redshift of the antipode) is also larger for~$a>0$. The CMB
anisotropy goes as~$|z_1|\sim|a|$, which suggests that it is unaffected by
the sign of~$a$.\footnote{This is not quite true, however, because it
depends on the time at which the surface of last scattering occurs
(i.e.,~on~$\chi\cmb$). This will be later for models with~$a>0$ (since the
epoch of observation is later, for the same~$H_0$), and the universe will
be more inhomogeneous. However, this effect turns out to be rather weak
and does not affect the conclusions we reach.}

However, $a>0$ implies that the universe will open up at $\ti=(1-b)/2a$, after
which time the universe is infinite in spatial extent (note that models
with~$a>0$ expand forever). Spatial infinity occurs at some
finite~$\psi=\psi_{\scriptscriptstyle{max}}$, where $W=0$ in~(\ref{confW}).
Applying the Copernican principle in this case becomes difficult~-- at least
in terms of producing exclusion diagrams. We will not give further attention
to this here.

\section{Conclusions}
\label{conclusions}

We have studied the physical, geometrical and observational characteristics
of a two-parameter family of inhomogeneous, perfect fluid cosmological models
which form a subclass of the models admitting isotropic radiation fields
found in~\cite{cla-bar99} and coincide with the model~I spacetimes
of~\cite{dab95}. We have shown that these models do not suffer from particle
horizons.

The inhomogeneity of these D\c{a}browski models makes the investigation
of their observational characteristics from any position essential, and
the simple conformal geometry of the models was used in
Secs.~\ref{Stephani}-\ref{obs-constr} to derive exact, analytic expressions
for the observational relations and other properties for any observer
position; the constraints imposed by observations were examined. For any model
parameters and any observer position we fixed the \emph{coordinate} time~$T$
of the present epoch (i.e.,~the time of observation) by demanding that the
value of~$H_0$ at that~$T$, given by~(\ref{Hubble-par}), was consistent
with current estimates of the Hubble constant~(\ref{min-H0}). We then
proceeded to test the consistency of the models with a variety of
observational constraints, the most important being age and the anisotropy
of the CMB.

Obviously, in any viable cosmological model the age of the universe must be
greater than the estimated age of of any of its components by the time the
expansion has slowed to the rate we measure today. This is position
dependent in our models: on a given surface of constant coordinate time
(determined by the expansion rate~$H_0$) the observers at the centre will
generally be older than their counterparts elsewhere (see
Fig.~\ref{exclusion-age}). As with all the tests we apply, we show the
constraints for the `worst case' observer position, i.e.,~ that which
excludes the largest region of parameter space.

Despite the fact that the D\c{a}browski models \emph{admit} an isotropic
radiation field \cite{cla-bar99}, this will not in general correspond
exactly to the actual cosmic background radiation field after decoupling.
This is because, although the models are homogeneous at early times (conformal
factor~$W=1$), the inhomogeneity ($W=W(\chi,t)$) that develops up to the
time of decoupling will leave its imprint in the CMB. In other words,
the EGS-type theorems prove the existence of an isotropic radiation field,
but whether this is realised in a particular spacetime depends on the physics
of decoupling. We used a pragmatic definition of the time of decoupling to
estimate the effect of the inhomogeneity on the CMB. The anisotropies in
the observed CMB temperature go as $1/(1+z)\sim1/W$: that is, all anisotropies
arise entirely from the inhomogeneities in the conformal factor $W$
\emph{at the time of decoupling}; the conformal flatness of the spacetimes
allows the light from the CMB surface to travel unmolested by any subsequent
deviation from homogeneity, and we see the CMB as a reflection of the
universe's inhomogeneity at that time. Since $t$ is small at
decoupling $W\sim1$ and the inhomogeneities are very small. The dipole
moment of the CMB is the largest moment, which we restrict to be smaller
than 10\% of the observed dipole, on the basis that present data cannot
show that Local Group motion and the CMB dipole agree to within
10\%~\cite{cla-rau99}.

In addition to these observational constraints, we required that the first
zero of the distance-redshift relations does not occur too close to the
observer, on the grounds that this would probably have been observed.
This ruled out models with small~$b$, which turned out to be fairly
homogeneous anyway.

The matter content of these models is unusual, in that there is no equation of
state. However, we know from \cite{cla-bar99} (see also \cite{thesis,sus99})
that they do admit a thermodynamic interpretation (i.e.,~definitions of number
density, temperature and entropy throughout spacetime consistent with the
first law of thermodynamics). We have shown that many, though not all, of the
D\c{a}browski models satisfy the weak, strong and dominant energy conditions
(Sec.~\ref{energy-cond}), so that their matter content cannot be ruled out as
obviously unphysical. We restricted our detailed analysis to the models that
satisfy all three energy conditions (the strong energy condition then
requires~$a<0$). Of course, we are not forced to accept the energy conditions.
Although they seem physically very reasonable conditions to impose on any
form of matter there are many examples in cosmology of matter that does not
satisfy them (quantum fields,for example, can exhibit negative energy
density). Given that quintessence models
\cite{framp99,lid99,coble-97,lid-sch98}, which produce an accelerating
expansion in FLRW models (consistent with the supernova data), must
break the strong energy condition, rejection of the $a>0$~models may be
premature, and we have considered their properties briefly in Sec.~\ref{agt0}
(see also~\cite{thesis}).

Our studies have shown that there is a significant subset of the D\c{a}browski
models that are markedly inhomogeneous but cannot be excluded on the basis of
the tests considered here. It is possible, \emph{for every observer} in
each of the models in the allowed regions of Figs. \ref{exclusion-H50}
and~\ref{exclusion-H70}, to choose the epoch of observation so that the
observed value of~$H_0$ is reproduced, the age is greater than the measured
age of the universe and there are no obviously unacceptable features at low
redshift ($z\lesssim 5$) in the observational relations. Most importantly,
though, the dipole in the cosmic background radiation would be considerably
smaller than the observed CMB dipole: despite the inhomogeneity of the models
the anisotropy they produce is very small. The fact that this is true for
every observer means that it is not possible to reject these models by
appealing to the Copernican principle. As a result, the standard assumption
that the observed high degree of isotropy about us combined with the
Copernican principle necessarily forces the universe to be homogeneous
(i.e.,~the cosmological principle) is seriously undermined.

\section*{Acknowledgments}

The authors would like to thank Martin Hendry and Mariusz D\c{a}browski for
their help and advice, and for initially suggesting the investigation of
Stephani models as plausible cosmological models. We would also like to thank
Roy Maartens, Bruce Bassett and Stephane Rauzy for help, advice and
encouragement. CAC was funded by a PPARC studentship.

\appendix

\section{Transformation to a Non-Central Position}
\label{chitrans}

We want to transform the metric~(\ref{metric-W}) from the $(\chi,\theta,\phi)$
coordinate system, whose origin is at the centre, to coordinates centred instead
on some observer at~$\chi=\psi$, while preserving the form of the \FRW\ part of
the metric. It is therefore necessary to identify the transformations of the
(homogeneous) \FRW\ spatial sections that leave the \FRW\ metric invariant,
i.e.,~the isometries of the spatial sections. This is simple. Since the spatial
sections of an \FRW\ model with positive curvature constant ($\Delta>0$) are
3-spheres, the isometries we require are 4-dimensional rotations (i.e.,~elements
of $SO(4)$, the isometry group of the 3-sphere).

A sphere of radius~$R$ in 4-dimensional space with cartesian coordinates
$(x,y,z,u)$ is defined by
\[
                x^2+y^2+z^2+u^2 = R^2.
\]
We have three coordinates on this sphere: $\chi$ and the two spherical polar
angles $\theta$ and~$\phi$. These are related to the cartesian coordinates by
\ba
       x & = & R\sin\chi \sin\theta \cos\phi \label{xofchi}\\
       y & = & R\sin\chi \sin\theta \sin\phi \\
       z & = & R\sin\chi \cos\theta \label{zofchi}\\
       u & = & R\cos\chi \label{uofchi}
\ea
The origin,~$\chi=0$, is then at $x=y=z=0$,~$u=R$. We are only interested in
4-rotations that move the origin, and, as the initial metric is spherically
symmetric (really spherically symmetric, not just conformally: even the
conformal factor is spherically symmetric about the centre), we need only
consider moving the observer in one direction, which we choose to be the
$z$~direction (i.e.,~to a position with non-zero~$z$, but~$x=y=0$). Clearly,
then, we are looking for a rotation in the $u-z$~plane. Since we have the
conformal factor as a function of~$\chi$ we want to find $\chi$ as a function of
the new coordinates. Starting with coordinates $\chi'$, $\theta'$ and~$\phi'$,
centred on some position~$\chi=\psi$, along with their primed cartesian
counterparts $x'$, $y'$, $z'$ and $u'$ (which are related in the same way as the
unprimed coordinates in (\ref{xofchi})--(\ref{uofchi})), a rotation back to the
original coordinates is given, in cartesian coordinates, by $x=x'$, $y=y'$ and
\ba
            z & = & \cos\psi z' + \sin\psi u', \nonumber\\
            u & = & -\sin\psi z' + \cos\psi u'. \label{urot}
\ea
(Note that at the origin of the primed coordinates, where $z'=0$ and~$u'=R$, we
have $u=R\cos\psi$, showing that~$\chi=\psi$ there, as required.)
Equation~(\ref{urot}), along with the primed versions of (\ref{zofchi})
and~(\ref{uofchi}), then immediately gives
\be
        \cos\chi = \cos\psi \cos\chi' - \sin\psi \sin\chi' \cos\theta',
\label{newchi}
\ee
and this is all we will need, since the only spatial coordinate that enters into
the original metric~(\ref{metric-W}) is~$\chi$, and that enters only
as~$\cos\chi$ ($2\sin^2\frac{\chi}{2}=1-\cos\chi$).

\section{Redshift in Conformally Related Spacetimes}
\label{appen}

We present two derivations of the redshift formula for spacetimes sharing some
of the simple properties of the D\c{a}browski models. The first can be used in
spacetimes that are conformal to simpler metrics for which the geodesics and
redshifts can be found. The second is valid for spacetimes that are conformal to
a spherically symmetric spacetime.

If we have two conformally related metrics, $g_{ab}=\Omega^2\bar{g}_{ab}$
($\Omega>0$), their associated metric connections are related by (see appendix~D
of \cite{wald})
\be
    \nabla_b V^a = \bar{\nabla}_b V^a + (\bar{\nabla}_b \ln\Omega) \,V^a
               + (V^c\bar{\nabla}_c \ln\Omega) \,\delta^a_b
               - (\bar{\nabla}_d \ln\Omega) \,g^{ad}g_{bc} V^c,
\label{connect}
\ee
for any vector field~$V^a$. Null geodesics with respect to~$g$ (or, more
correctly, with respect to~$\nabla$) satisfy $g_{ab}k^a k^b=0$ and
\be
                     k^b \nabla_b k^a = 0,
\label{geodesic}
\ee
where~$k^a$ is the tangent vector to the geodesic. Applying~(\ref{connect})
gives
\[
0 = k^b\nabla_b k^a  =  k^b\bar{\nabla}_b k^a + 2
k^b\bar{\nabla}_b\ln\Omega\,k^a
        - g^{ad}\bar{\nabla}_d\ln\Omega\,\left(g_{bc}k^b k^c\right)
               =  \frac{1}{\Omega^4} (\Omega^2 k^b)\bar{\nabla}_b
              (\Omega^2 k^a),
\]
and we see immediately that
\be
                  \bar{k}^a = \Omega^2 k^a
\label{kbar}
\ee
is the tangent vector to a null geodesic with respect to~$\bar{\nabla}$. So,
every null geodesic of~$\nabla$ corresponds to a null geodesic of~$\bar{\nabla}$
(and vice versa, since we can repeat the above steps interchanging $g$
and~$\bar{g}$ and putting~$\Omega\mapsto 1/\Omega$). If the geodesics with
respect to $\bar{\nabla}$ are known explicitly then we can find them easily for
$\nabla$.

To find the redshift, though, we also need the velocities of emitter and
observer. A four-velocity satisfies $g_{ab}u^au^b = -c^2$, and if we define
\be
                 \bar{u}^a = \Omega u^a
\label{ubar}
\ee
then $\bar{u}$ is a four-velocity with respect to~$\bar{g}$:
$\bar{g}_{ab}\bar{u}^a\bar{u}^b=-c^2$. Redshift is calculated from the ratio of
the emitted frequency~$\nu_{\scriptscriptstyle{E}} =
\left.u_ak^a\right|_{\scriptscriptstyle{E}}$ to the observed
frequency~$\nu_{\scriptscriptstyle{O}} =
\left.u_ak^a\right|_{\scriptscriptstyle{O}}$. Using (\ref{kbar})
and~(\ref{ubar}) we have
\[
 u_ak^a = g_{ab}u^ak^b = \Omega^2\bar{g}_{ab}\frac{\bar{u}^a}{\Omega}
            \frac{\bar{k}^b}{\Omega^2}
        = \frac{1}{\Omega}\bar{u}_a\bar{k}^a,
\]
which means that
\be
1+z=\frac{\left. u_{a}k^{a}\right|_{\scriptscriptstyle{E}}}
{\left.u_{b}k^{b}\right|_{\scriptscriptstyle{O}}}=
\frac{\Omega_{\scriptscriptstyle{O}}}{\Omega_{\scriptscriptstyle{E}}}
\frac{\left.\bar{u}_{a}\bar{k}^{a}\right|_{\scriptscriptstyle{E}}}
{\left.\bar{u}_{b}\bar{k}^{b}\right|_{\scriptscriptstyle{O}}}=
\frac{\Omega_{\scriptscriptstyle{O}}}{\Omega_{\scriptscriptstyle{E}}}
(1+\bar{z}).
\label{redshift1}
\ee
where $\bar{z}$ is the redshift associated with $\bar{g}_{ab}$ for the
fundamental velocity~$\bar{u}$. If the paths of null rays in the
spacetime~$\bar{g}_{ab}$ and the redshift formula for the velocity~$\bar{u}$ are
known then~(\ref{redshift}) gives the redshift in the true spacetime. For the
D\c{a}browski models $\bar{g}_{ab}$ is an \FRW\ metric, the conformal factor is
$\Omega=1/W$ (see~(\ref{metric-W})) and $\bar{u}$ is the usual \FRW\ comoving
velocity field. The well-known expression for redshift in \FRW\ spacetimes,
$1+\bar{z} = R_{\scriptscriptstyle{O}}/R{\scriptscriptstyle{E}}$, then gives
\be
          1+z = \frac{R_{\scriptscriptstyle{O}}}{W_{\scriptscriptstyle{O}}}
                \frac{W_{\scriptscriptstyle{E}}}{R_{\scriptscriptstyle{E}}}.
\label{redshift}
\ee

When the true spacetime is conformal to a spherically symmetric spacetime the
radial null geodesics connecting any point with an observer at the centre are
obviously purely radial (since their paths are not affected by the conformal
factor). They are therefore given (in terms of coordinates $r$ and~$t$ with
respect to which the spherical symmetry is manifest) by some
function~$t_{\scriptscriptstyle{O}}(r_{\scriptscriptstyle{E}},
t_{\scriptscriptstyle{E}})$ relating the time,~$t_{\scriptscriptstyle{O}}$ ,
that the light ray is received by the observer, to the time of
emission,~$t_{\scriptscriptstyle{E}}$, for an object at
radius~$r_{\scriptscriptstyle{E}}$. This is just the lookback-time relation.
Redshift, as the ratio of proper time intervals~$d\tau_{\scriptscriptstyle{O}}$
at the observer to proper time intervals~$d\tau_{\scriptscriptstyle{E}}$ at the
emitter, is then given by
\be
    1+z \equiv \frac{d\tau_{\scriptscriptstyle{O}}}
    {d\tau_{\scriptscriptstyle{E}}}
           =  \frac{d\tau_{\scriptscriptstyle{O}}}{dt_{\scriptscriptstyle{O}}}
             \frac{dt_{\scriptscriptstyle{O}}}{d\tau_{\scriptscriptstyle{E}}}
            = \frac{d\tau_{\scriptscriptstyle{O}}}{dt_{\scriptscriptstyle{O}}}
              \left(\frac{\partial t_{\scriptscriptstyle{O}}}
              {\partial r_{\scriptscriptstyle{E}}}
                    \frac{dr_{\scriptscriptstyle{E}}}
                    {d\tau_{\scriptscriptstyle{E}}}
               +    \frac{\partial t_{\scriptscriptstyle{O}}}
               {\partial t_{\scriptscriptstyle{E}}}
                    \frac{dt_{\scriptscriptstyle{E}}}
                    {d\tau_{\scriptscriptstyle{E}}}\right)
            =  \frac{1}{u^t_{\scriptscriptstyle{O}}}
                 \left(\frac{\partial t_{\scriptscriptstyle{O}}}
                 {\partial r_{\scriptscriptstyle{E}}}
                    u^r_{\scriptscriptstyle{E}}
               +    \frac{\partial t_{\scriptscriptstyle{O}}}
               {\partial t_{\scriptscriptstyle{E}}}
                    u^t_{\scriptscriptstyle{E}}\right).
\label{zDtdef}
\ee
(When the coordinates $r$ and~$t$ are comoving~-- $u^r=0$~-- the $r$-derivative
term disappears.) This will provide an analytic expression for the redshift
whenever the lookback-time equation can be integrated. For the D\c{a}browski
models:
\be
          u^\chi=0,  \hskip 1.0cm   u^\ti = \frac{c}{|g_{00}|^{1/2}} = W
\label{ut}
\ee
and the lookback time can be derived directly from the metric: on the past null
cone of the observer $ds=0=d\theta=d\phi$, leading to an expression
for~$d\chi/d\ti$, which, when integrated, gives
\[
              \chi = c\sqrt{\Delta}\int_\ti^\T(\chi,\ti) \frac{d\ti'}{R(\ti')}.
\]
Differentiating this with respect to~$\ti$ at fixed~$\chi$ then gives
\[
     \frac{\partial t_{\scriptscriptstyle{O}}}
     {\partial t_{\scriptscriptstyle{E}}}
         \equiv    \frac{\partial \T}{\partial \ti} =
                \frac{R_{\scriptscriptstyle{O}}}{R_{\scriptscriptstyle{E}}},
\]
which, together with (\ref{zDtdef}) and~(\ref{ut}), results in the
expression~(\ref{redshift}) for the redshift.


\begin{thebibliography}{10}

\bibitem{cla-bar99}
C.~A. Clarkson and R.~K. Barrett, Class. Quantum Grav. {\bf 16},  3781  (1999).

\bibitem{ellis75}
G.~F.~R. Ellis, Q. J. R. Astron. Soc. {\bf 16},  245  (1975).

\bibitem{goo}
J.~Goodman, Phys. Rev. D, {\bf 52}, 1821 (1995).

\bibitem{ehlers-68}
J. Ehlers, P. Geren, and R.~K. Sachs, J.~Math. Phys. {\bf 9},  1344  (1968).

\bibitem{stoeg-95}
W.~R. Stoeger, R. Maartens, and G.~F.~R. Ellis, Astrophys.~J. {\bf 443},  1
  (1995).

\bibitem{thesis}
C.~A. Clarkson, Ph.D. Thesis. The University of Glasgow, 1999.
 {\it astro-ph/0008089}

\bibitem{kant-sachs66}
R. Kantowski and R.~K. Sachs, J.~Math. Phys. {\bf 7},  443  (1966).

\bibitem{ell98}
G.~F.~R. Ellis, gr-qc/9812046v3.

\bibitem{nils-99}
U.~S. Nilsson, C. Uggla, J. Wainwright, and W.~C. Lim, Astrophys.~J. Lett.
   {\bf 521}, L1  (1999).

\bibitem{lem}
G. Lema\^\i tre, Ann. Soc. Scient. Bruxelles A, {\bf 53}, 51 (1933).

\bibitem{tolm34}
R.~C. Tolman, Proc. Nat. Acad. Sci. USA {\bf 20}, 169   (1934).

\bibitem{bond47}
H. Bondi, Mon. Not. R. Astron. Soc. {\bf 107}, 410 (1947).

\bibitem{hel-lak-I84}
C. Hellaby and K. Lake, Astrophys.~J. {\bf 282},  1  (1984).

\bibitem{hel-lak85}
C. Hellaby and K. Lake, Astrophys.~J. {\bf 290},  381  (1985).

\bibitem{rindler-sus89}
W. Rindler and D. Suson, Astron. Astrophys. {\bf 218},  15  (1989).

\bibitem{goi-mm97}
L.~J. Goicoechea and J.~M. Martin-Mirones, Astron. Astrophys.
    {\bf 186},  22  (1987).

\bibitem{maar-95}
R. Maartens, N.~P. Humphreys, D.~R. Matravers, and W.~R. Stoeger, Class.
  Quantum Grav. {\bf 13},  253  (1996).

\bibitem{cel99}
M.~C\'el\'erier.
\newblock {\em astro-ph/9907206}, 1999.

\bibitem{cel-sch98}
M.~C\'el\'erier and J.~Schneider.
\newblock {\em Phys. Lett. A}, 249:37--45", 1998.

\bibitem{mus}
N. Mustapha, C.~Hellaby and G.~F.~R. Ellis, Mon. Not. R. Astron. Soc.
{\bf 292}, 817 (1997).

\bibitem{tom96}
K. Tomita, Astrophys~J. {\bf 461},  507  (1996).

\bibitem{tomita95}
K. Tomita, Astrophys.~J. {\bf 451},  1  (1995).

\bibitem{mof-tat95}
J.~W. Moffat and D.~C. Tatarski, Astrophys.~J. {\bf 453},  17  (1995).

\bibitem{nakao-95}
K. Nakao {\it et~al.}, Astron.~J. {\bf 453},  541  (1995).

\bibitem{kras98}
A. Krasi\'{n}ski, gr-qc/9806039.

\bibitem{humph-97}
N.~P. Humphreys, R. Maartens, and D.~R. Matravers, Astrophys.~J. {\bf 477},
  47   (1997).

\bibitem{step67a}
H. Stephani, Commun. Math. Phys. {\bf 4},  137  (1967a).

\bibitem{step67b}
H. Stephani, Commun. Math. Phys. {\bf 5},  337  (1967b).

\bibitem{kram-80}
D. Kramer {\it et~al.}, {\em Exact Solutions of Einstein's Field Equations}
  (Cambridge University Press, Cambridge, England, 1980).

\bibitem{Kras83}
A. Krasi\'{n}ski, Gen. Relativ. Gravit. {\bf 15},  673  (1983).

\bibitem{kras97}
A. Krasi\'{n}ski, {\em Inhomogeneous Cosmological Models}
  (Cambridge University Press, Cambridge, England, 1997).

\bibitem{dab-hen98}
M.~P. D\c{a}browski, and M.~A, Hendry, Astrophys.~J. {\bf 498},  67  (1998).

\bibitem{perl-97}
S. {Perlmutter} {\it et~al.}, Astrophys.~J. {\bf 483},  565  (1997).

\bibitem{dab95}
M.~P. D\c{a}browski, Astrophys.~J. {\bf 447},  43  (1995).

\bibitem{bar-cl99}
R.~K. Barrett and C.~A. Clarkson (to be submitted).

\bibitem{perl-99}
S. {Perlmutter} {\it et~al.}, Astrophys.~J. {\bf 517},  565  (1999).

\bibitem{bon-col88}
C. Bona and B. Coll, Gen. Relativ. Gravit. {\bf 20},  297  (1988).

\bibitem{sus99}
R.~A. Sussman, gr-qc/9908019.

\bibitem{lorpet86}
D. Lorenz-Petzold, Astrophys. Astron. {\bf 7},  155  (1986).

\bibitem{wald}
R.~M. Wald, {\em General Relativity} (The University of Chicago Press,
  Chicago,  1984).

\bibitem{hawk-ell}
S.~W. Hawking and G.~F.~R. Ellis, {\em The Large Scale Structure of
  Space-Time} (Cambridge University Press, Cambridge, England, 1973).

\bibitem{kraus98}
L.~M. Krauss, Astrophys.~J. {\bf 501},  461  (1998).

\bibitem{peeb98}
P.~J.~E. Peebles, astro-ph/9810497.

\bibitem{framp99}
P.~H. Frampton, astro-ph/9901013.

\bibitem{lid99}
A.~R. Liddle, astro-ph/9901041.

\bibitem{coble-97}
K. Coble, S. Dodelson, and J.~A. Frieman, Phys. Rev.~D {\bf 55},  1851
  (1997).

\bibitem{lid-sch98}
A.~R. Liddle and R.~J. Scherrer, astro-ph/9809272.

\bibitem{gol-ellis98}
M. Goliath and G.~F.~R. Ellis, gr-qc/9811068.

\bibitem{ellis-78}
G.~F.~R. Ellis, R. Maartens, and S.~D. Nel, Mon. Not. R. Astron. Soc.
  {\bf 184}, 439 (1978).

\bibitem{cla-bar99b}
C.~A. Clarkson and R.~K. Barrett, to be submitted to Class. Quantum Grav.
  (1999b).

\bibitem{dab93}
M.~P. D\c{a}browski, J.~Math. Phys. {\bf 34},  1447  (1993).

\bibitem{Vilenkin}
A. Vilenkin, Phys.~Rev.~Lett. {\bf 46},  1169  (1981).

\bibitem{dab-stel89}
M.~P. \dab\ and J. Stelmach, Astron.~J. {\bf 97},  978  (1989).

\bibitem{kris-sac66}
J. Kristian and R.~K. Sachs, Astrophys.~J. {\bf 143},  379  (1966).

\bibitem{Mac-Ellis-II}
M.~A.~H. MacCallum and G.~F.~R. Ellis, Commun. Math. Phys. {\bf 19},  31
   (1970).

\bibitem{pet-lee96}
V. Petrosian and T.~T. Lee, Astrophys. J.~Lett.  {\bf 467}  L29  (1996).

\bibitem{cow-90}
L.~L. Cowie {\it et~al.}, Astrophys. J.~Lett. {\bf 360},  L1  (1990).

\bibitem{lilly-91}
S.~J. {Lilly}, L.~L. {Cowie}, and J.~P. {Gardner}, Astrophys.~J. {\bf 369},
  79   (1991).

\bibitem{chab-97}
B. Chaboyer, P. Demarque, P.~J. Kernan, and L.~M. Krauss, Astrophys.~J.
  {\bf 494}  96  (1998).

\bibitem{bert-97}
E. Bertschinger {\it et~al.},  in {\em The Evolution of the Universe}, edited
  by G. B\"{o}rner and  S. G\"{o}ttlober (John Wiley \& Sons,
  Chichester, 1997),
  Chap.~17.

\bibitem{madau96}
P. Madau {\it et~al.}, Mon. Not. R. Astron. Soc. {\bf 283},  1388  (1996).

\bibitem{sadat98}
R. Sadat, A. Blanchard, B. Guiderdoni, and J. Silk, Astron. Astrophys.
    {\bf 331} L69 (1998).

\bibitem{loeb99}
A. Loeb, astro-ph/9907187.

\bibitem{wang98}
Y. {Wang}, N. {Bahcall}, and E.~L. {Turner}, Astrophys.~J. {\bf 116},  2081
  (1998).

\bibitem{saw-97}
M. J. {Sawicki}, H. {Lin} and H. K. C. {Yee}, Astrophys. J. {\bf 113}, 1 (1997).

\bibitem{cooray99}
A.~R. {Cooray}, J.~M. {Quashnock}, and M.~C. {Miller}, Astrophys.~J.
   {\bf 511}, 562  (1999).

\bibitem{part97}
B. Partridge,  in {\em From Quantum Fluctuations to Cosmological Structures},
  edited by D. Valls-Gabaud, M.~A. Hendry, P. Molaro, and K. Chamcham
  (ASP Conf. Ser.~126, 1997), pp.141--184.

\bibitem{lynd-bell88}
D. Lynden-Bell. et~al., Astrophys.~J. {\bf 326},  19  (1988).

\bibitem{schmoldt-99}
I. Schmoldt {\it et~al.}, astro-ph/9901087.

\bibitem{ries-95}
A.~G. Riess, W.~H. Press, and R.~P. Kirshner, Astrophys.~J. {\bf 445},
   (1995).

\bibitem{lau-pos91}
T.~R. Lauer and M. Postman, Astrophys.~J. {\bf 425},  418  (1991).

\bibitem{cla-rau99}
C.~A. Clarkson, S. Rauzy, and R.~K. Barrett, (in preparation).

\bibitem{rub-76a}
V.~C. {Rubin}, M.~S. {Roberts}, J.~A. {Graham}, J. {Ford}, W.~K., and N.
  {Thonnard}, Astrophys.~J. {\bf 81},  687  (1976a).

\bibitem{rub-76b}
V.~C. {Rubin}, M.~S. {Roberts}, N. {Thonnard}, and J. {Ford}, W.~K.,
    Astrophys.~J. {\bf 81},  719  (1976b).

\bibitem{jam-91}
P.~A. {James}, R.~D. {Joseph}, and C.~A. {Collins}, Mon. Not. R. Astron. Soc.
  {\bf 248},  444  (1991).

\bibitem{hamuy-96}
M. Hamuy {\it et~al.}, Astron.~J. {\bf 112},  2391  (1996).

\end{thebibliography}

\end{document}